\documentstyle[amssymb,prd,aps,epsfig]{revtex}

\begin{document}
%
%
\def\be{\begin{equation}}
\def\ee{\end{equation}}
\def\RR{{\Bbb{R}}}
\def \wg#1{\mbox{\boldmath ${#1}$}}
\newfont{\ssg}{cmssbx10}
\def \w#1{\mbox{\ssg {#1}}}
\draft
\title{Binary black holes in circular orbits.
I. A global spacetime approach}
\author{Eric Gourgoulhon\footnote{e-mail: {\tt Eric.Gourgoulhon@obspm.fr}},
Philippe Grandcl\'ement\footnote{present address: 
Department of Physics and Astronomy, Northwestern University, 
Evanston, IL 60208, USA;
e-mail: {\tt PGrandclement@northwestern.edu}}
and Silvano Bonazzola\footnote{e-mail: {\tt Silvano.Bonazzola@obspm.fr}} }
\address{D\'epartement d'Astrophysique Relativiste et de Cosmologie \\
  UMR 8629 du C.N.R.S., Observatoire de Paris, \\
  F-92195 Meudon Cedex, France
}
\date{11 October 2001}
\maketitle

\begin{abstract}
We present a new approach to the problem of binary black holes in the 
pre-coalescence stage, i.e. when the notion of orbit has still some meaning. 
Contrary to previous numerical treatments which 
are based on the initial value formulation of general relativity on a 
(3-dimensional) spacelike hypersurface, our approach deals with the full
(4-dimensional) spacetime. This permits a rigorous definition of the orbital 
angular velocity. Neglecting the gravitational radiation reaction, 
we assume that the
black holes move on closed circular orbits, which amounts to endowing the
spacetime with a helical Killing vector. We discuss the choice of
the spacetime manifold, the desired properties of the 
spacetime metric, as well as the choice
of the rotation state for the black holes. As a simplifying assumption,
the space 3-metric is approximated by a conformally flat one.
The problem is then reduced to solving five of the
ten Einstein equations, which are derived here, as well as the boundary 
conditions on the black hole surfaces and at spatial infinity.
We exhibit the remaining five Einstein equations and propose to use
them to evaluate the error induced by the conformal flatness approximation. 
The orbital angular velocity of the system is computed through
a requirement which reduces to the classical virial theorem at the
Newtonian limit.  
\end{abstract}


\begin{center}
We dedicate this work to the memory of our friend and collaborator 
Jean-Alain Marck.
\end{center}

\section{Background and motivation} \label{s:intro}

Binary black holes have been the subject of numerous studies
in the past two decades, both from the analytical and
numerical point of view. These studies are motivated by the
fact that the coalescence of two black holes is expected to
be one of the strongest sources of gravitational waves
detectable by the interferometric detectors LIGO, GEO600, TAMA300
and VIRGO, currently coming on-line \cite{GrishLPPS01}.

From the analytical point of view, the most recent works
are based on the post-Newtonian formalism (see e.g. Ref.~\cite{Blanc00}
for a review) or on the effective one-body approach developed
by Buonanno and Damour \cite{BuonaD99,BuonaD00,Damou01}. In these
works, the black holes are treated as point mass particles\footnote{Note
however the (approximate) analytical  solution derived by Alvi
\cite{Alvi00} by matching a post-Newtonian metric to two perturbed
Schwarzschild metrics},
which is a very good approximation when the black holes are
far apart. For closer configurations, one may turn instead to
some numerical approach. The numerical studies can be divided in two
classes: (i) the initial value problem for two black holes
(see Ref.~\cite{Cook00} for a review) and (ii) the time evolution
of the initial data (see Ref.~\cite{Seide99} for a review and
Refs.~\cite{BrandCGHLL00,AlcubBBLNST01,BakerBCLT01} for recent results).
One of the major problems in this respect is to get physically relevant
initial data. Indeed, initial data representing two black holes
have been obtained long ago by Misner \cite{Misne63} and Lindquist
\cite{Lindq63}, as well as Brill and Lindquist \cite{BrillL63} (a
modern discussion of these solutions can be found in
Ref.~\cite{Giuli98} or Appendices A and B of Ref.~\cite{AndraP97}). 
However these solutions correspond to
two momentarily {\em static} black holes and are therefore far
from representing some stage in the evolution of an isolated binary black hole 
in our universe.

Based on the seminal work of Bowen and York
\cite{BowenY80}, Kulkarni, Shepley and York \cite{KulkaSY83}
have described a procedure to get initial data representing 
binary black hole with arbitrary positions, masses, spins and 
momenta. This procedure, known as {\em conformal imaging},
has been used by Cook and other authors
\cite{Cook91,CookCDKMO93,Cook94,PfeifTC00,DieneJKN00}
to numerically construct 3-dimensional spacelike
hypersurfaces representing these initial data. These solutions 
constitute some generalization to the non-static regime
of the Misner-Lindquist solution
\cite{Misne63,Lindq63}, the spacelike hypersurface having the 
same topology: two isometric asymptotically-flat sheets connected
by two Einstein-Rosen bridges. 
Also based on the Bowen and York's work \cite{BowenY80}, another
approach, the so-called {\em puncture method},
has been undertaken by Brandt and Br\"ugmann \cite{BrandB97}
and used recently by Baumgarte \cite{Baumg00}. 
The resulting solutions also have arbitrary positions, masses, spins and 
momenta but their topology is different from that of the conformal-imaging
approach: the spacelike hypersurface has now {\em three}
asymptotically-flat sheets, which are not isometric. These three sheets
are connected among themselves by two Einstein-Rosen bridges; one of
the sheets contains two throats and is supposed to represent our
universe. These solutions hence constitute some generalization to the
non-static case of the Brill-Lindquist solution \cite{BrillL63}.
Both the conformal-imaging and the puncture approaches assume 
that the metric of the spacelike hypersurface is
conformally flat. A third approach to the problem relaxes this assumption;
it has been developed recently by Matzner, Maroronetti and collaborators
\cite{MatznHS99,MarroHLLMS00,MarroM00}. These authors use
a linear combination of two boosted Kerr-Schild metrics as the
conformal 3-metric in York's treatment of initial conditions
\cite{York79}. 

The main drawback of the three approaches described above is that
they contain freely specifiable parameters, related to the values of positions,
momenta and spins of the black holes, and it is difficult to figure
out which parameters correspond to a physical configuration.
In particular, it is not obvious at all how to select, among all these
configurations, those that correspond to binary black holes on closed
circular orbits. Of course, the circular orbits are approximate
representations of the exact orbital motion, which is inspiralling
due to the loss of energy and angular momentum via gravitational
radiation. However in the regime where the radiation reaction
time scale is much longer than the orbital time scale, i.e.
before the last stable orbit, we expect that the binary system can be 
approximated by a sequence of tighter and tighter circular
orbits. Note that the gravitational radiation reaction makes initially
elliptic orbits become circular \cite{Peter64}. 
It is thus legitimate to search for such orbits.
This problem has been addressed by Cook \cite{Cook94} and
Pfeiffer et al. \cite{PfeifTC00} within the
conformal-imaging approach and by Baumgarte \cite{Baumg00}
within the puncture approach\footnote{To our knowledge, the Kerr-Schild
approach has not been used yet to get circular orbits, the article
\cite{MarroM00} providing results only for black holes in
hyperbolic motions.} (see e.g. \cite{Baumg01} for a review of these
computations). Although they differ in the topology of the spacelike
hypersurface (two-sheeted for \cite{Cook94,PfeifTC00}
against three-sheeted for \cite{Baumg00}), both sets of studies rely
on the {\em effective-potential method} proposed by Cook \cite{Cook94}.
This method amounts to defining the binding energy by a somewhat ad-hoc
formula, and to define the angular velocity of a circular orbit
by minimizing the binding energy with respect to the angular momentum
at fixed total energy and separation. This method can be criticized
on the following ground: the only well defined global quantities
on an asymptotically flat spacelike hypersurface are the ADM total mass $M$
and ADM total linear momentum. The latter can be chosen to be zero without
any loss of generality. With some restrictions on the asymptotic gauge,
the total angular momentum $J$ can also be defined \cite{York80}.
Defining the binding energy would require the notion
of individual mass for each hole, let say $M_1$ and $M_2$,
in order to set $E_{\rm bind} = M - M_1 - M_2$. 
However there does not exist any
unique definition of the individual masses $M_1$ and $M_2$ for a 
binary black hole. In Refs.~\cite{Cook94,PfeifTC00,Baumg00} the authors
define $M_1$ and $M_2$ via the formula which 
relates the mass of a Kerr black hole to its horizon area 
and its angular momentum.
However such a formula is strictly demonstrated only for a isolated rotating
black hole (Kerr spacetime). In particular, it does not take into
account any tidal effect. 

Our approach for finding circular orbits of binary black holes 
is very different: instead of considering 3-dimensional spacelike hypersurfaces,
we adopt from the very beginning a 4-dimensional point of view, i.e. 
we consider a full spacetime containing two moving black holes. Of course, 
in order to make the problem tractable, we introduce some approximations,
the most significant being the assumption of strictly circular orbits, 
which amounts to endowing our spacetime with a Killing vector field 
(helical symmetry). An interesting pay-off is that the
orbital angular velocity $\Omega$ can be
unambiguously defined as the rotation rate of the Killing field with respect
to some asymptotically inertial observers. This definition does not suffer
from the ambiguity of the 3-dimensional approaches and is made possible
only because we have re-introduced {\em time} in the problem. 

The presentation of this new approach is organized as follows. We set up the problem
in Sec.~\ref{s:formulation}, starting 
from the explicit construction of the spacetime manifold, introducing
a metric, as well as a corresponding isometry on it,
and finally imposing the helical symmetry. The Einstein equations are
then considered in Sec.~\ref{s:einstein}, first in their general form
and then after the assumption of a conformally flat 3-metric. 
Global quantities such as the ADM mass and the total angular momentum
are discussed in this section, as well as the virial prescription for
the orbital velocity. Section~\ref{s:asymptotic} deals with the
asymptotic behavior of the shift vector and the extrinsic curvature
tensor and discusses the connection between helical symmetry and asymptotic
flatness. Finally Sec.~\ref{s:conclu} contains the concluding remarks.

\section{Formulation} \label{s:formulation}

\subsection{Spacetime manifold}

\subsubsection{Construction}  \label{s:construct}

\begin{figure}
\centerline{ \epsfig{figure=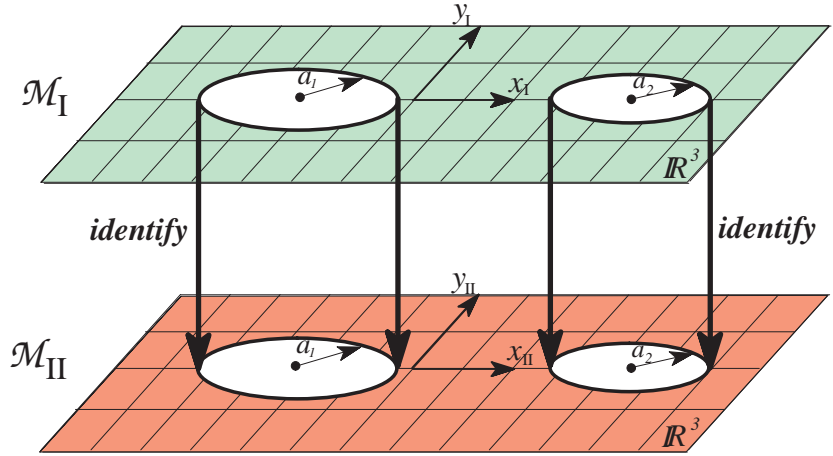,height=6cm} }
\caption[]{\label{f:manifold} 
Construction of the spacetime manifold.}  
\end{figure}

We consider the spacetime to be a differentiable manifold $\cal M$ with
the topology of the real line $\RR$ times
the Misner-Lindquist manifold \cite{Misne63,Lindq63}.
More precisely, for any couple of positive numbers $(a_1,a_2)$ and
any couple of real numbers $(x_1,x_2)$ such that $|x_1-x_2| > a_1 + a_2$, 
let us consider the subset of ${\RR^3}$ obtained by removing the
interior of balls of radius $a_1$ and $a_2$ and center $x=x_1$ and
$x=x_2$:
\be
	{\cal E} := \left\{ (x,y,z) \in {\RR^3},
		(x-x_1)^2 + y^2 + z^2 \geq a_1^2 \ \mbox{and} \ 
		(x-x_2)^2 + y^2 + z^2 \geq a_2^2 \right\} \ . 
\ee
Let us call $S_1$ and $S_2$ the 2-spheres defining
the ``inner'' boundaries of $\cal E$:
\be
	 S_1 := \left\{ (x,y,z) \in \RR^3,
		(x-x_1)^2 + y^2 + z^2 =  a_1^2  \right\} \ ,	
\ee
\be
	 S_2 := \left\{ (x,y,z) \in \RR^3,
		(x-x_2)^2 + y^2 + z^2 =  a_2^2  \right\} \ . 	
\ee
Let us consider two copies ${\cal E}_{\rm I}$ and ${\cal E}_{\rm II}$
of $\cal E$ and define ${\cal M}_{\rm I} := \RR \times {\cal E}_{\rm I}$
and ${\cal M}_{\rm II} := \RR \times {\cal E}_{\rm II}$. The spacetime
manifold $\cal M$ is be then defined as the union 
${\cal M}_{\rm I} \cup {\cal M}_{\rm II}$ with 
both $S_1$ and $S_2$ of each copy identified
(see Fig.~\ref{f:manifold}). 
The reader is referred to Sect.~IV of Ref.~\cite{Misne63} or Sect.~II
of Ref.~\cite{Lindq63} for a precise construction of the manifold 
structure in the vicinity of $S_1$ and $S_2$. 
The part ${\cal M}_{\rm I}$ of $\cal M$ will be designed hereafter as
the {\em upper space} and the part ${\cal M}_{\rm II}$ as the 
{\em lower space}. 
The boundaries ${\cal S}_1:=\RR \times S_1$ and ${\cal S}_2:=\RR \times S_2$ 
between ${\cal M}_{\rm I}$ and ${\cal M}_{\rm II}$ are called respectively
{\em throat 1} and {\em throat 2}. 

Hereafter we label by $(t,x_{\rm I},y_{\rm I},z_{\rm I})$
the points of ${\cal M}_{\rm I}$ considered as a part of
$\RR\times\RR^3$ (${\cal E}_{\rm I}$ being a part of $\RR^3$), 
and by $(t,x_{\rm II},y_{\rm II},z_{\rm II})$
the points of ${\cal M}_{\rm II}$ considered as a part of
$\RR\times\RR^3$. The corresponding two charts will be called the
{\em canonical coordinate systems}. These two charts cover 
$\cal M$ minus the two throats.  
The whole manifold $\cal M$ can be covered entirely 
by a single coordinate system:
\be
	\begin{array}{cccl}
		C_1 : & {\cal M} & \longrightarrow & \RR^4 \\
	& P & \longmapsto & \left\{ \begin{array}{ll}
	(t,x_{\rm I},y_{\rm I},z_{\rm I}) & {\rm if\ } 
					P \in {\cal M}_{\rm I}  \\
  {\cal I}_1(t,x_{\rm II},y_{\rm II},z_{\rm II}) & {\rm if\ }
					P \in {\cal M}_{\rm II}  
			\end{array} \right.
	\end{array} \ , 
\ee
where ${\cal I}_1 : \RR^4 \mapsto \RR^4$ denotes the inversion
through the 2-sphere $S_1$ : 
\be
	{\cal I}_1(t,x,y,z) = \left( t,\ 
	{a_1^2 (x-x_1) \over (x-x_1)^2 + y^2 +z^2} + x_1 ,\ 
	{a_1^2 y \over (x-x_1)^2 + y^2 +z^2}, \ 
	{a_1^2 z \over (x-x_1)^2 + y^2 +z^2} 	\right) \ . 
\ee
In a similar way, one can introduce the 
coordinate system $C_2$ associated with throat 2.

\begin{figure}
\centerline{ \epsfig{figure=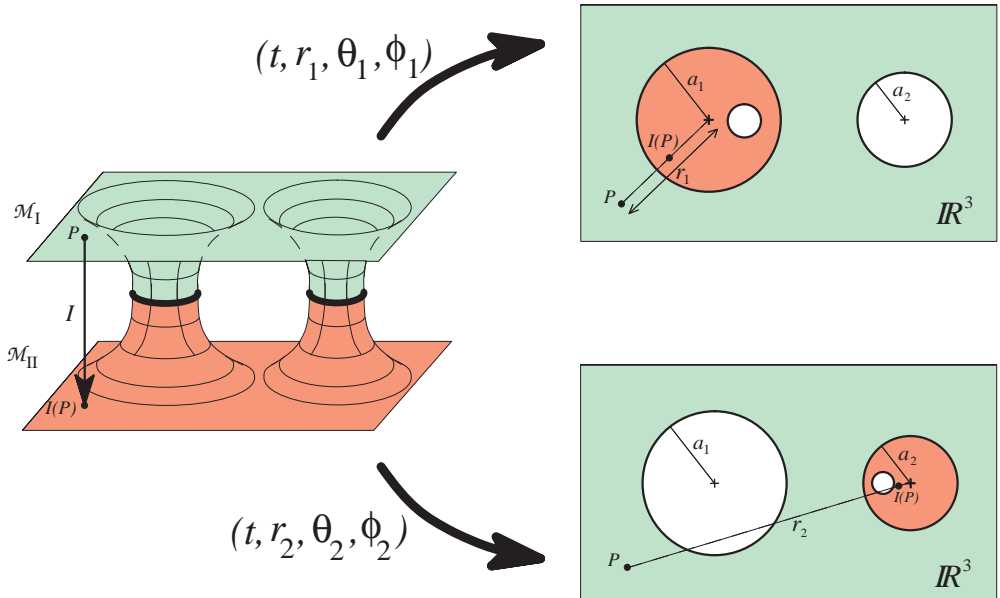,height=8cm} }
\caption[]{\label{f:manicoord} 
Coordinate systems $(t,r_1,\theta_1,\varphi_1)$ and 
$(t,r_2,\theta_2,\varphi_2)$
on the spacetime manifold $\cal M$. 
Shown here is a $t={\rm const}$ section of $\cal M$, with the dimension 
in the $\theta$ direction suppressed, leaving only $(r_1,\varphi_1)$
or $(r_2,\varphi_2)$.}
\end{figure}

In the coordinate system $C_1$ or $C_2$, the throats are not located 
at constant coordinate values. Therefore, it is more convenient
to introduce instead the polar coordinate
system $(t,r_1,\theta_1,\varphi_1)$ centered on throat 1, as follows:
\begin{equation}
  \mbox{for}\ P \in {\cal M}_{\rm I}, \quad  
	\left\{ \begin{array}{l}
		x_{\rm I} = r_1 \sin\theta_1 \cos\varphi_1 + x_1 \\
		y_{\rm I} = r_1 \sin\theta_1 \sin\varphi_1 \\
		z_{\rm I} = r_1 \cos\theta_1 
	\end{array} \right.
	\qquad (a_1 \leq r_1 < + \infty)
\end{equation}
and
\begin{equation}
  \mbox{for}\ P \in {\cal M}_{\rm II}, \quad  
	\left\{ \begin{array}{l}
		x_{\rm II} = {a_1^2\over r_1} \sin\theta_1 \cos\varphi_1 + x_1 \\
		y_{\rm II} = {a_1^2\over r_1} \sin\theta_1 \sin\varphi_1 \\
		z_{\rm II} = {a_1^2\over r_1} \cos\theta_1 
	\end{array} \right.
	\qquad (0<  r_1 \leq a_1) \ . 
\end{equation}
The throat 1 corresponds to $r_1 = a_1$. 
The polar coordinate system $(t,r_2,\theta_2,\varphi_2)$ centered on
throat 2 is introduced similarly.
Note that any of the two coordinate
systems $(t,r_1,\theta_1,\varphi_1)$ and $(t,r_2,\theta_2,\varphi_2)$ 
covers the whole spacetime manifold $\cal M$. For the
$(t,r_1,\theta_1,\varphi_1)$ system,
$\cal M_{\rm I}$ corresponds to $a_1\leq r_1 < +\infty$ 
and $\cal M_{\rm II}$ to $0<r_1 \leq a_1$. 
Similarly, for $(t,r_2,\theta_2,\varphi_2)$ system, 
$\cal M_{\rm I}$ corresponds to $a_2\leq r_2 < +\infty$  and $\cal M_{\rm II}$
to $0< r_2 \leq a_2$ (see Fig.~\ref{f:manicoord}).

\subsubsection{Canonical mapping}

From the very construction of $\cal M$, we have at our disposal the
canonical mapping (see Fig.~\ref{f:manifold})
\be \label{e:def_I}
	\begin{array}{lccc}
	I : & {\cal M}_{\rm I} & \longrightarrow & {\cal M}_{\rm II} \\
	 & (t,x_{\rm I},y_{\rm I},z_{\rm I}) & \longmapsto & 
	   (t = t, x_{\rm II} = x_{\rm I},
		y_{\rm II} = y_{\rm I}, z_{\rm II} = z_{\rm I})		
	\end{array} 
\ee   
Note that in terms of the $(t,r_1,\theta_1,\varphi_1)$ coordinate system, 
this map can be written as an inversion through the sphere $r_1=a_1$
(see Fig.~\ref{f:manicoord}):
\be \label{e:I(r1)}
	I(t,r_1,\theta_1,\varphi_1) = \left( t, {a_1^2\over r_1},
				\theta_1, \varphi_1 \right) \ . 
\ee   
In terms of the $(t,r_2,\theta_2,\varphi_2)$ coordinate system, it looks
like an inversion as well:
\be \label{e:I(r2)}
	I(t,r_2,\theta_2,\varphi_2) = \left( t, {a_2^2\over r_2},
				\theta_2, \varphi_2 \right) \ . 
\ee   

Let $(x^\alpha)$ be a coordinate system on ${\cal M}_{\rm I}$
[for instance $(x^\alpha)=(t,x_{\rm I},y_{\rm I},z_{\rm I})$,
$(x^\alpha)=(t,r_1,\theta_1,\varphi_1)$ or 
$(x^\alpha)=(t,r_2,\theta_2,\varphi_2)$] and 
$(y^\mu)$ be a coordinate system on ${\cal M}_{\rm II}$
[for instance $(y^\mu)=(t,x_{\rm II},y_{\rm II},z_{\rm II})$,
$(y^\mu)=(t,r_1,\theta_1,\varphi_1)$ or 
$(y^\mu)=(t,r_2,\theta_2,\varphi_2)$]. The map $I$ is fully characterized
by its components $I^\mu$ with respect to the coordinates $(x^\alpha)$
and $(y^\mu)$: the image $I(P)$ of a point $P\in{\cal M}_{\rm I}$ with
coordinates $(x^0,x^1,x^2,x^3)$ has the coordinates
\be
	y^\mu = I^\mu(x^0,x^1,x^2,x^3) \ . 
\ee

Let us now examine the action of the map $I$ on various fields
on $\cal M$. 
If $f$ is a scalar field on ${\cal M}_{\rm II}$, $I$ induces a scalar field on
${\cal M}_{\rm I}$ through 
\be \label{e:def_I_fonct}
	I_* f := f \circ I \ , 
\ee
i.e. 
\be
	I_*f[P] = f[I(P)] \ , 
\ee
for any point $P$ of ${\cal M}_{\rm I}$. 

Also $I$ maps any vector field $\w{v}$ on ${\cal M}_{\rm I}$ to a vector
field $I_*\w{v}$ on ${\cal M}_{\rm II}$ through 
\be \label{e:def_I_vect}
	I_*\w{v}(f) := \w{v}(I_* f) \ , 
\ee
for any scalar field $f$ on ${\cal M}_{\rm II}$. If vectors are
represented by their components with respect to the coordinate
bases $\partial/\partial x^\alpha$ and $\partial/\partial y^\mu$, one
has
\be
	I_*\w{v}(f) = (I_*\w{v})^\mu {\partial f \over \partial y^\mu}
\ee
and, according to definitions (\ref{e:def_I_vect}) and (\ref{e:def_I_fonct}),
\be
	I_*\w{v}(f) = v^\alpha {\partial \over \partial x^\alpha} 
			f\left( I^\mu(x^\beta) \right)
		      = v^\alpha {\partial f \over \partial y^\mu}
				{\partial I^\mu \over \partial x^\alpha} \ .
\ee
Hence the matrix of the mapping $I_*$ between vectors on
${\cal M}_{\rm I}$ and vectors on ${\cal M}_{\rm II}$ is given by
the Jacobian matrix of $I$:
\be \label{e:I_vect_jacob}
	(I_*\w{v})^\mu [I(P)] = {\partial I^\mu \over \partial x^\alpha}
				\, v^\alpha [P] \ , 
\ee
where $P$ denotes any point of ${\cal M}_{\rm I}$. 

The action of $I$ on vectors can be used to define the action of $I$
on 1-forms as follows:
$I$ maps any 1-form $\wg{\omega}$ on ${\cal M}_{\rm II}$
to a 1-form $I_*\wg{\omega}$ on ${\cal M}_{\rm I}$ through
\be \label{e:def_I_form}
   I_*\wg{\omega}(\w{v}) := \wg{\omega}(I_*\w{v}) \ , 
\ee
for any vector field $\w{v}$ on ${\cal M}_{\rm I}$. 
If 1-forms are represented by their components with respect to the coordinate
bases $dx^\alpha$ and $dy^\mu$, one has
\be \label{e:I_form_1}
	I_*\wg{\omega}(\w{v}) = 
		(I_*\wg{\omega})_\alpha dx^\alpha(\w{v})
		= (I_*\wg{\omega})_\alpha v^\alpha
\ee
and, according to definition (\ref{e:def_I_form}),
\be \label{e:I_form_2}
	I_*\wg{\omega}(\w{v}) = \omega_\mu dy^\mu(I_*\w{v})
 		      = \omega_\mu (I_*\w{v})^\mu
		      = \omega_\mu {\partial I^\mu \over \partial x^\alpha}
				v^\alpha \ , 
\ee
where the third equality arises from Eq.~(\ref{e:I_vect_jacob}). 
Comparing Eqs.~(\ref{e:I_form_1}) and (\ref{e:I_form_2}) leads to
\be
	(I_*\wg{\omega})_\alpha [P] = 
		{\partial I^\mu \over \partial x^\alpha} \, \omega_\mu[I(P)]  
\ee
at any point $P$ in ${\cal M}_{\rm I}$.

Similarly, the action of $I$ on bilinear forms can be defined as follows:
$I$ associates any bilinear form $\w{T}$ on ${\cal M}_{\rm II}$ to
a bilinear form $I_*\w{T}$ on ${\cal M}_{\rm I}$ according to
\be \label{e:def_I_bilin_comp}
	I_*\w{T}(\w{v},\w{w}) := \w{T}(I_*\w{v},I_*\w{w}) \ . 
\ee
One can show easily that in terms of the components with respect to
the coordinate bases $\w{d}x^\alpha \otimes \w{d}x^\beta$ and
$\w{d}y^\mu \otimes \w{d}y^\nu$,
\be \label{e:I_bilin_comp}
	(I_*\w{T})_{\alpha\beta} [P] = {\partial I^\mu \over \partial x^\alpha}
				   {\partial I^\nu \over \partial x^\beta}
				   \, T_{\mu\nu}[I(P)] 
\ee
at any point $P$ in ${\cal M}_{\rm I}$.

\subsection{Spacetime metric} \label{s:spacetime_metric}

\subsubsection{Properties} \label{s:metric_prop}

We endow $\cal M$ with a Lorentzian metric $\w{g}$ 
with the following properties:

(1) $\w{g}$ is asymptotically flat at the ends of ${\cal M}_{\rm I}$
	and ${\cal M}_{\rm II}$ : 
\begin{eqnarray}
   \lim_{x_{\rm I}^2+y_{\rm I}^2+z_{\rm I}^2\rightarrow\infty} 
			\w{g} & = & \wg{\eta} \ , \label{e:asymp_flat1} \\
   \lim_{x_{\rm II}^2+y_{\rm II}^2+z_{\rm II}^2\rightarrow\infty} 
			\w{g} & = & \wg{\eta} \ , \label{e:asymp_flat2}
\end{eqnarray}
where $\wg{\eta}$ is a flat metric.

(2) the canonical mapping $I$ is an isometry of $\w{g}$:
\be \label{e:isometrie}
	I_*\w{g} = \w{g} \ . 
\ee

(3) The $t={\rm const}$ sections of $\cal M$ are maximal 
spacelike hypersurfaces with respect to $\w{g}$. 

The assumption (1) is introduced because we consider only isolated
systems. Its connection with the quasi-stationarity hypothesis will
be discussed in Secs.~\ref{s:hel_killing} and \ref{s:non_flat}.

The assumption (2) is motivated by the fact that the Schwarzschild
and Kerr spacetimes possess such an isometry. This can be readily seen
when using isotropic (quasi-isotropic for Kerr) coordinates instead
of the standard Schwarzschild (Boyer-Lindquist) ones. 
By virtue of Eq.~(\ref{e:I_bilin_comp}), the isometry condition 
(\ref{e:isometrie}) can be 
expressed in terms of the components of $\w{g}$ at any point $P$ in
${\cal M}_{\rm I}$ :
\be  \label{e:isometrie_g_cov}
	{\partial I^\mu \over \partial x^\alpha}
				   {\partial I^\nu \over \partial x^\beta}
				   \, g_{\mu\nu}[I(P)] 
	= g_{\alpha\beta}[P] \ . 
\ee
It is also useful to write the isometry condition on the contravariant
components of the metric tensor; by means of a generalization
of Eq.~(\ref{e:I_vect_jacob}), one gets:
\be \label{e:isometrie_g_contra}
	g^{\mu\nu} [I(P)] = {\partial I^\mu \over \partial x^\alpha}
			    {\partial I^\nu \over \partial x^\beta}
				\, g^{\alpha\beta} [P] \ .  
\ee

The assumption (3) is motivated by the well-known good properties 
of maximal slicing \cite{York79,SmarrY78a}, among which there is
the singularity avoidance.

\subsubsection{3+1 decomposition}

In this article we use the standard 3+1 formalism of general relativity
\cite{York79}, foliating the spacetime by a family of spacelike
hypersurfaces.  
From the very construction of $\cal M$, a natural foliation
is by the $t=const$ hypersurfaces $\Sigma_t$, where $t$ is the same
coordinate as that introduced above. By virtue of assumption (3), 
this constitutes a maximal slicing of spacetime. 

Let us denote by $\w{n}$ the future directed unit normal to $\Sigma_t$.
Being normal to $\Sigma_t$, $\w{n}$ should be collinear to the gradient
of $t$:
\be \label{e:def_lapse}
	\w{n} = - N \wg{\nabla} t \ , 
\ee
where $N$ is the {\em lapse function}, which can be seen as a normalization
factor such to ensure that $\w{n}\cdot\w{n} := \w{g}(\w{n},\w{n}) = -1$.
Let us now examine the behavior of $\w{n}$ under the isometry $I$. 
By the definition (\ref{e:def_I}), $I$ preserves the hypersurface $\Sigma_t$. 
According to the definition (\ref{e:def_I_bilin_comp}),
the square of the norm of $I_*\w{n}$ is 
\be
	\w{g}(I_*\w{n},I_*\w{n})
			       = (I_*\w{g})(\w{n},\w{n}) \ .
\ee
But thanks to the isometry condition $I_*\w{g} = \w{g}$, the last term
in this equation is simply $\w{n}\cdot\w{n}=-1$.
Hence
\be
	I_*\w{n}\cdot I_*\w{n} = - 1 \ ,
\ee 
i.e. $I$ preserves the norm of $\w{n}$.
Similarly, for any vector $\w{v}$ tangent to $\Sigma_t$, $I$ preserves
the scalar product $\w{n}\cdot\w{v} = 0$, so that 
$I_*\w{n}\cdot I_*\w{v} = 0$. But since $\Sigma_t$ is globally 
invariant under $I$, $I_*\w{v}$ represents any vector tangent to $\Sigma_t$,
so that $I_*\w{n}$ is in fact normal to $\Sigma_t$. Having the same norm
than $\w{n}$, we conclude that
\be \label{e:I_n}
	I_*\w{n} = \pm \w{n} \ . 
\ee
Since $t$, considered as a scalar field on $\cal M$, is preserved by $I$,
so is its gradient and the relation (\ref{e:def_lapse}), combined with
(\ref{e:I_n}) results then in the following transformation law for the
lapse function:
\be \label{e:I_lapse}
	I_* N = \pm N \ . 
\ee

\begin{figure}
\centerline{ \epsfig{figure=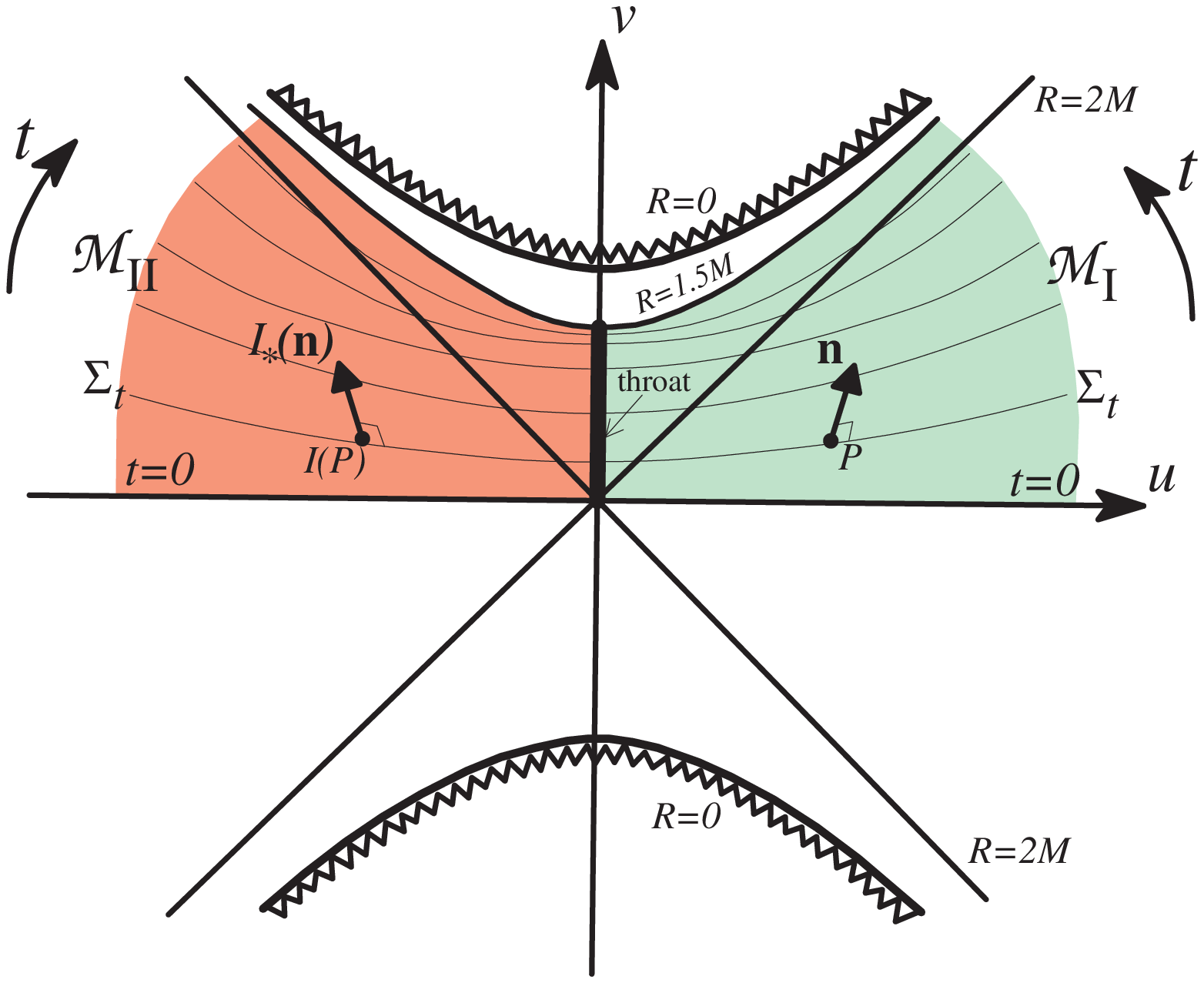,height=7cm} \quad 
	     \epsfig{figure=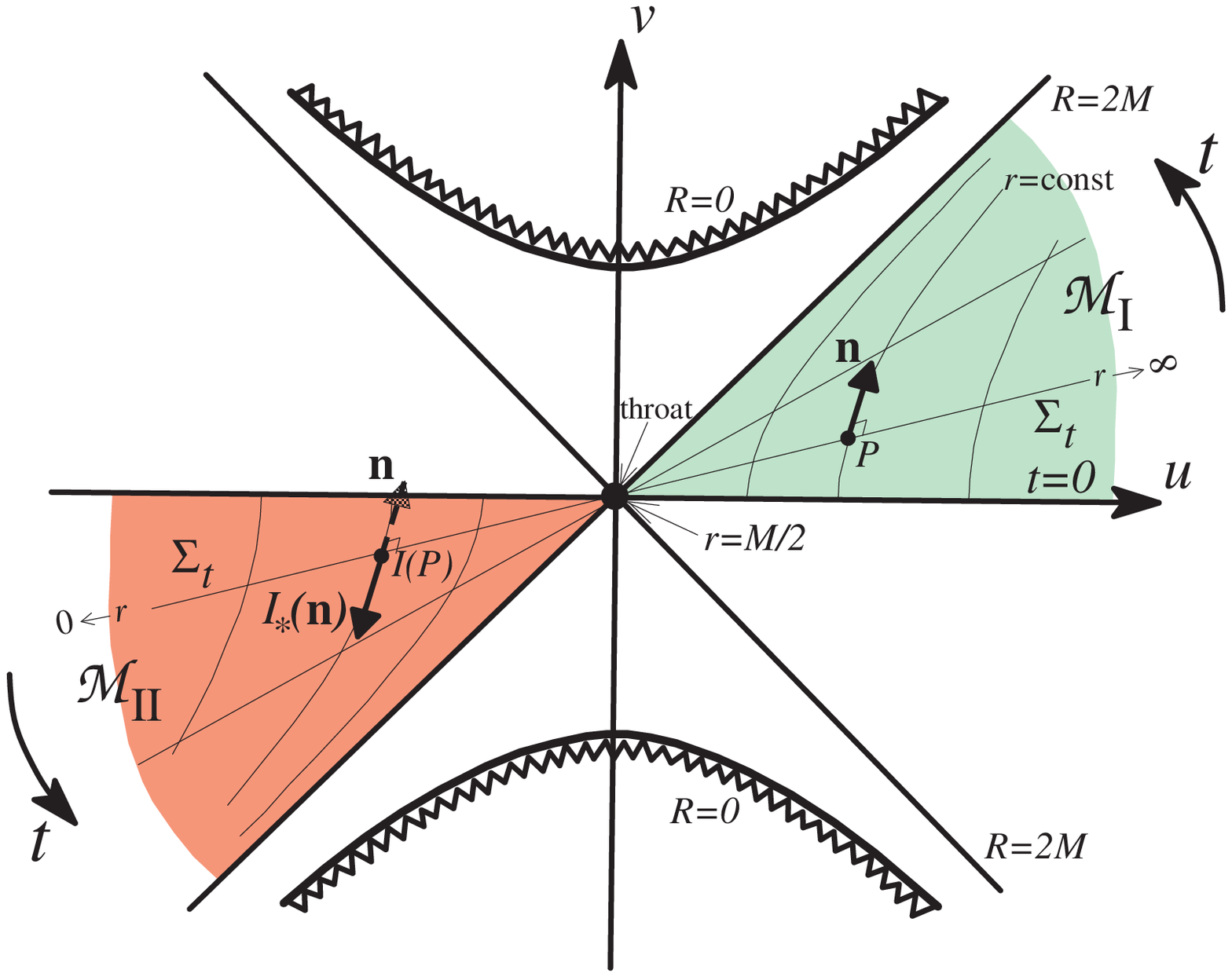,height=7cm} }
\vspace{1cm}
\caption[]{\label{f:kruskal} 
Kruskal diagrams showing the slicing of the extended Schwarzschild spacetime
by two families of maximal hypersurfaces; {\em left:} lapse function symmetric
with respect to the isometry $I$; {\em right:} lapse function antisymmetric 
with respect to the isometry $I$. $R$ denotes the standard Schwarzschild
radial coordinate and $r$ the isotropic one.}  
\end{figure}

In order to understand the significance of the $\pm$ sign in Eqs.~(\ref{e:I_n})
and (\ref{e:I_lapse}), let us consider the case of a single static black
hole, i.e. the (extended) Schwarzschild spacetime. Two kinds of
maximal slicing of this spacetime are depicted in a Kruskal diagram
in Fig.~\ref{f:kruskal}, starting from the same initial hypersurface
$v=0,\ t=0$. The first one corresponds to a symmetric
lapse [sign $+$ in Eqs.~(\ref{e:I_lapse}) and (\ref{e:I_n})].
The throat is located at $u=0$; the slicing penetrates under the event 
horizon ($R=2M$), and accumulates on the spacelike
hypersurface $R=1.5M$
\cite{EstabWCDST73,BeigO98}. The second slicing corresponds to an antisymmetric
lapse [sign $-$ in Eqs.~(\ref{e:I_lapse}) and (\ref{e:I_n})]. In fact,
it corresponds to the standard Schwarzschild solution in isotropic 
coordinates:
\be \label{e:ds2_Schwarz}
	ds^2 = - N^2 dt^2 + \left(1 + {M\over 2r} \right) ^4
		\left[ dr^2 + r^2(d\theta^2 + \sin^2\theta d\varphi^2)
		\right] \ , 
\ee
with
\be \label{e:N_Schwarz}
	N = { 1 - M/2r \over 1 + M/2r } \ . 
\ee
This lapse function is clearly antisymmetric under 
the transformation $I: r\mapsto M^2/(4 r)$ across the throat
located at $r=M/2$. The negative value of the lapse for $r<M/2$ is
easily understandable when looking to Fig.~\ref{f:kruskal}: while $t$
is running upward on the right part of the diagram (corresponding to
${\cal M}_{\rm I}$), it is running downward on the left
part (corresponding to ${\cal M}_{\rm II}$). Since in the Kruskal
diagram the future direction is everywhere upward, the lapse should
be negative in ${\cal M}_{\rm II}$.

Let us now consider a coordinate system $(x^i)$
on each $\Sigma_t$.
For instance, it can be chosen in one of 
the three coordinate atlas introduced so far:
$\{(x_{\rm I},y_{\rm I},z_{\rm I}),\, (x_{\rm II},y_{\rm II},z_{\rm II})\}$,
$\{(r_1,\theta_1,\varphi_1)\}$ and $\{(r_2,\theta_2,\varphi_2)\}$.
$(t,x^i)$ constitutes then a coordinate system on $\cal M$.
The {\em shift vector} $\wg{\beta}$ associated with the coordinates
$(t,x^i)$ is defined by the following orthogonal split of the
coordinate basis vector $\partial/\partial t$:
\be \label{e:def_beta}
	{\partial \over\partial t} = N \w{n} + \wg{\beta}
	\qquad \mbox{with} \qquad \w{n}\cdot\wg{\beta} = 0 \ .
\ee
Since the transformation $I$ is purely spatial
$\partial/\partial t$ is preserved by it. By virtue of Eqs.~(\ref{e:I_n})
and (\ref{e:I_lapse}), the product $N \w{n}$ is also invariant with
respect to $I$. Consequently
\be \label{e:I_beta}
	I_*\wg{\beta} = \wg{\beta} \ . 
\ee

The 3-metric induced by $\w{g}$ on the hypersurfaces $\Sigma_t$ is
\be
	\wg{\gamma} = \w{g} + \w{n}\otimes\w{n} \ . 	
\ee
From Eqs.~(\ref{e:isometrie}) and (\ref{e:I_n}), we obtain immediately
that $I$ is also an isometry for the 3-metric $\wg{\gamma}$:
\be \label{e:I_gamma}
	I_*\wg{\gamma} = \wg{\gamma} \ . 
\ee

The components of the metric tensor can be expressed in terms of
the lapse function and the components of the shift vector and the
3-metric, according to the classical formula
\be \label{e:metric_3p1}
   g_{\mu\nu} dx^\mu dx^\nu = -(N^2 - \beta_i \beta^i) \, dt^2
		+ 2\beta_i \, dt \, dx^i + \gamma_{ij} \, dx^i \, dx^j \ .  
\ee

The extrinsic curvature tensor $\w{K}$ of the hypersurface $\Sigma_t$
is given by the Lie derivative of the 3-metric along the flow defined 
by the normal to $\Sigma_t$: 
\be \label{e:def_extr_curv}
	\w{K} = -{1\over 2} \pounds_{\w{n}} \wg{\gamma} \ . 
\ee
By the symmetry properties (\ref{e:I_n}) and (\ref{e:I_gamma}), we
obtain that 
\be \label{e:I_K}
	I_*\w{K} = \pm \w{K} \ .
\ee

\subsubsection{Explicit isometry conditions in polar coordinates}

In what follows, we consider only polar coordinate systems centered
on one of the two throats, i.e. either the system
$(t,r_1,\theta_1,\varphi_1)$ introduced in Sec.~\ref{s:construct}
or $(t,r_2,\theta_2,\varphi_2)$.
For the sake of clarity we will drop the indices $1$ or $2$ on
$r$, $\theta$ and $\varphi$. It should be understood that the formulas
will be valid for either coordinate system. 
The Jacobian matrix of $I$ with respect to $(t,r,\theta,\varphi)$
is easily deduced from Eq.~(\ref{e:I(r1)}) or (\ref{e:I(r2)}):
\be
  {\partial I^\mu \over \partial x^\alpha} = \mbox{diag}
	\left( 1,\ -{a^2\over r^2},\ 1,\ 1 \right) \ , 
\ee
where $a$ denotes either $a_1$ or $a_2$. 
From the isometry condition (\ref{e:isometrie_g_contra}) expressed
on $g^{tt}$, we get, since $g^{tt} = -1/N^2$ : 
\be \label{e:isom_expl_N}
	N[I(P)]^2 = N[P]^2  \ , 
\ee
for any point point $P$ in ${\cal M}_{\rm I}$, i.e. we recover
the already established relation (\ref{e:I_lapse}). 
From the isometry condition (\ref{e:isometrie_g_contra}) expressed
on $g^{ti}$ we get
\begin{eqnarray}
   \beta^r[I(P)] & = & - {a^2\over r^2} \beta^r[P] \label{e:I_beta_r} \\
   \beta^\theta[I(P)] & = & \beta^\theta[P] 	\label{e:I_beta_t} \\
   \beta^\varphi[I(P)] & = & \beta^\varphi[P] \ , \label{e:I_beta_p}
\end{eqnarray} 
where we have used $g^{ti} = \beta^i/N^2$ and Eq.~(\ref{e:isom_expl_N})
to go from $g^{ti}$ to $\beta^i$. Again note that we recover
the isometry condition (\ref{e:I_beta}).

Finally the isometry condition (\ref{e:isometrie_g_cov}) expressed
on $g_{ij}=\gamma_{ij}$ results in 
\begin{eqnarray}
  \gamma_{rr}[P] & = & {a^4\over r^4} \gamma_{rr}[I(P)] \label{e:iso_grr}\\
  \gamma_{r\theta}[P] & = & - {a^2\over r^2} \gamma_{r\theta}[I(P)]
  					\label{e:iso_grt}\\
  \gamma_{r\varphi}[P] & = & - {a^2\over r^2} \gamma_{r\varphi}[I(P)]
  					\label{e:iso_grp} \\
  \gamma_{\theta\theta}[P] & = & \gamma_{\theta\theta}[I(P)]
  					\label{e:iso_gtt}\\
  \gamma_{\theta\varphi}[P] & = & \gamma_{\theta\varphi}[I(P)]
  					\label{e:iso_gtp}\\
  \gamma_{\varphi\varphi}[P] & = & \gamma_{\varphi\varphi}[I(P)]
  					\label{e:iso_gpp} \ .
\end{eqnarray}
Comparing Eqs.~(\ref{e:isometrie}) and (\ref{e:I_K}), we see that 
the isometries properties of the components $K_{ij}$ of $\w{K}$ are
the same as those above for $\gamma_{ij}$, except possibly for an
opposite sign.

\subsubsection{Choice of the isometry sign}

As discussed above, the behavior of the foliation with respect to 
the isometry $I$ involves a $+$ or $-$ sign in the transformation
rules of the unit normal [Eq.~(\ref{e:I_n})], lapse function 
[Eq.~(\ref{e:I_lapse})] and extrinsic curvature [Eq.~(\ref{e:I_K})]. 
In this article, we choose the sign to be the minus one. This is motivated
by the fact that the maximal slicing with the $+$ sign of the 
Schwarzschild spacetime (left part of Fig.~\ref{f:kruskal})
does not respect the stationarity of the problem, i.e. 
the Killing vector $\partial/\partial t$ of Schwarzschild geometry 
does not carry a slice of that foliation into another slice
\cite{EstabWCDST73} 
(see also Sec.~IV of Ref.~\cite{BernsHS89}). 
On the contrary, the slicing with the $-$ sign
(right part of Fig.~\ref{f:kruskal}) respects the stationarity of the
problem. 
It seems to us more appealing to use a slicing which in the case of
a single black hole, makes the problem time-independent. We regard
the artificial time dependence resulting from the $+$ sign as
an unnecessary complication. Beside simplicity, another advantage of the
$-$ sign choice
is to allow us to test the numerical code by comparison with the standard
form of the Schwarzschild or Kerr metric in the special case of a single 
black hole. 

Thus, from now on, we set
\be \label{e:I_n_minus}
	I_*\w{n} = - \w{n} \ , 
\ee
\be \label{e:I_lapse_minus}
	I_* N = - N \ ,
\ee
and 
\be \label{e:I_K_minus}
	I_*\w{K} = - \w{K} \ .
\ee
Eq.~(\ref{e:I_lapse_minus}) can be explicited for any point 
$P$ in ${\cal M}_{\rm I}$:
\be \label{e:I_lapse_minus_expl}
	N[I(P)] = - N[P] \ , 
\ee
which amounts to choosing
the $-$ sign when taking the square root of Eq.~(\ref{e:isom_expl_N}). 

\subsubsection{Boundary conditions on the throats}

An immediate consequence of Eq.~(\ref{e:I_lapse_minus_expl}) is that the
lapse function vanishes on the two throats:
\be \label{e:lapse_zero_throat}
	N|_{{\cal S}_1} = 0 \qquad \mbox{and} \qquad N|_{{\cal S}_2} = 0 \ .
\ee
Indeed from the very definition of $I$ [Eq.~(\ref{e:def_I})] and the
construction of $\cal M$ by identifications of the two copies of
${\cal S}_1$ or ${\cal S}_2$, every point $P$ in ${\cal S}_1$ or 
${\cal S}_2$ is a fixed point for $I$. Hence Eq.~(\ref{e:I_lapse_minus_expl})
results in $N[P] = - N[P]$ on ${\cal S}_1$ and ${\cal S}_2$. 

Similarly, Eq.~(\ref{e:I_beta_r}) implies that the $r$ component
of the shift vector vanishes on the throats:
\be \label{e:shift_r_zero_throat}
	\left. \beta^{r_1} \right| _{{\cal S}_1} = 0 
	\qquad \mbox{and} \qquad
	\left. \beta^{r_2} \right| _{{\cal S}_2} = 0 \ .  
\ee
Taking the first derivatives of Eqs.~(\ref{e:I_beta_r})-(\ref{e:I_beta_p}),
we get the additional following relations on the throats, as a consequence
of the isometry of the shift vector:
\begin{eqnarray} 
  \left. {\partial \beta^r \over \partial \theta} \right| _{{\cal S}} 
						& = & 0 \label{e:dbr/dt_s}\\ 
  \left. {\partial \beta^r \over \partial \varphi} \right| _{{\cal S}} 
						& = & 0 \label{e:dbr/dp_s}\\ 
  \left. {\partial \beta^\theta \over \partial r} \right| _{{\cal S}} 
						& = & 0 \label{e:dbt/dr_s}\\ 
  \left. {\partial \beta^\varphi \over \partial r} \right| _{{\cal S}} 
						& = & 0 \label{e:dbp/dr_s} \ , 
\end{eqnarray}
where we have dropped the indices 1 or 2 on $r$, $\theta$, $\varphi$
and $\cal S$. Note that relations (\ref{e:dbr/dt_s}) and (\ref{e:dbr/dp_s})
could have been obtained also as consequences of 
Eq.~(\ref{e:shift_r_zero_throat}) since the throats are located at a
constant value of the coordinate $r$. 

Equations (\ref{e:iso_grr})-(\ref{e:iso_gpp})
and their first derivatives give the following values for the
3-metric on the throats:
\begin{eqnarray} 
& & \left. \left( {\partial \gamma_{rr}\over \partial r} 
	+ 2 {\gamma_{rr} \over r} \right) \right| _{{\cal S}} = 0 
					\label{e:isom_grr_der} \\
 & & \left. \gamma_{r\theta} \right| _{{\cal S}} = 0 
					\label{e:isom_grt} \\
 & & \left. {\partial \gamma_{r\theta} \over \partial \theta}  
					\right| _{{\cal S}} = 0 
	\qquad \mbox{and} \qquad
	\left. {\partial \gamma_{r\theta} \over \partial \varphi}  
					\right| _{{\cal S}} = 0 
					\label{e:isom_grt_der} \\
 & & \left. \gamma_{r\varphi} \right| _{{\cal S}} = 0 
					\label{e:isom_grp} \\
 & & \left. {\partial \gamma_{r\varphi} \over \partial \theta}  
					\right| _{{\cal S}} = 0 	
		\qquad \mbox{and} \qquad
 \left. {\partial \gamma_{r\varphi} \over \partial \varphi}  
					\right| _{{\cal S}} = 0 
						\label{e:isom_grp_der}\\
 & & \left. {\partial \gamma_{\theta\theta} \over \partial r} 
						\right| _{{\cal S}} = 0
						\label{e:isom_gtt_der} \\
 & & \left. {\partial \gamma_{\theta\varphi} \over \partial r} 
						\right| _{{\cal S}} = 0 
						\label{e:isom_gtp_der} \\
 & & \left. {\partial \gamma_{\varphi\varphi} \over \partial r} 
						\right| _{{\cal S}} = 0 
						\label{e:isom_gpp_der} \ . 
\end{eqnarray}

\subsubsection{Apparent horizons}

As a direct consequence of the isometry hypothesis, the throats
$S_1$ and $S_2$ are minimal 2-surfaces of the spatial hypersurface
$\Sigma_t$. 
Moreover, as shown by Cook \& York \cite{CookY90}, the fact that 
$\w{K}$ is antisymmetric with respect to the isometry $I$ 
[Eq.~(\ref{e:I_K_minus})] implies that $S_1$ and $S_2$ are apparent
horizons. 

\subsection{Quasi-stationarity hypothesis}

\subsubsection{Helical Killing vector} \label{s:hel_killing}

As discussed in Sect.~\ref{s:intro}, we consider binary black holes
in the quasi-steady stage, i.e. prior to any orbital instability, so
that the notion of closed circular orbits is meaningful. 
Following Detweiler \cite{Detwe89}, 
we translate these assumptions in terms of the spacetime geometry by
demanding that there exists a Killing vector field $\wg{\ell}$ 
such that, near spacelike infinity, 
\be \label{e:helical}
 \wg{\ell} \rightarrow  {\partial \over \partial t_0} + \Omega \,
	{\partial\over\partial \varphi_0} \ ,
\ee
where $t_0$ and $\varphi_0$ are respectively the time coordinate
and the azimuthal coordinate associated with an asymptotically 
inertial observer, and $\Omega$ is a constant, representing the orbital
angular velocity with respect to the asymptotically 
inertial observer. 
Let us call $\wg{\ell}$ the {\em helical Killing vector}. 
We refer the reader to \cite{FriedUS01} for a detailed description
of this concept.

The helical symmetry amounts to neglecting outgoing gravitational
radiation in the dynamics of spacetime. 
For non-axisymmetric systems --- as binaries are ---
it is well known that imposing $\wg{\ell}$ as an exact 
Killing vector leads to a spacetime which is not asymptotically flat
\cite{GibboS83}. In Sec.~\ref{s:non_flat}, we will exhibit explicitly
how the deviation from asymptotical flatness arises. 
However, from a physical point of view,
the exact helical symmetry is
too strong an assumption because it assumes that the binary system is rotating
on a fixed orbit since the past time infinity. Doing so, it has filled
the entire space with gravitational waves, such that their total energy
is a diverging quantity, whence the impossibility of asymptotic flatness.
A weaker assumption, which is compatible with asymptotic flatness and
sounds physically more reasonable, is the following one. Due to the
reaction to gravitational radiation the binary system is in fact spiraling.
Therefore in the past time infinity, it was infinitely separated.
As a consequence, the amount of emitted gravitational waves was very
weak. The integral of their energy density is now a converging 
quantity, which allows for asymptotic flatness. The quasi-stationarity 
hypothesis should then be understood as imposing a helical
Killing vector on a part of spacetime {\em limited in time}.

It is natural to demand that the isometry associated with the
Killing vector $\wg{\ell}$ preserves, not only $({\cal M},\w{g})$ as a whole,
but also the sub-structure of $\cal M$ defined by ${\cal M}_{\rm I}$,
${\cal M}_{\rm II}$ and the two throats. This amounts to demanding that
for any of the coordinates system $(t,x^i)$ introduced above, where
$t$ is the coordinate used explicitly in the construction of $\cal M$,
\be \label{e:l=d/dt}
	\wg{\ell} =  {\partial \over \partial t} \ . 
\ee
The above equality means that $t$ is an ignorable coordinate. It does
not mean that the problem is stationary in the usual sense of this word,
for $\wg{\ell}$ is not timelike at spatial infinity: by virtue of relation
(\ref{e:helical}), 
$\wg{\ell}\cdot\wg{\ell} \sim \Omega^2 (x_{\rm I}^2 + y_{\rm I}^2) > 0$
when $x_{\rm I},y_{\rm I} \rightarrow \infty$.

\subsubsection{Rotation states of the black holes}

The above geometrical assumptions are intended to correspond
to a physical system of two black holes in a quasi-steady state. 
We have not specified yet the rotation state (spin)
of each black hole. In this article, we consider
{\em synchronized} (or {\em corotating}) black holes. 
This rotation state can be translated geometrically by
demanding that the two throats be {\em Killing horizons} \cite{Carte69}
associated with the helical symmetry. 
This means that each null-geodesic generator of ${\cal S}_1$
and ${\cal S}_2$ must be parallel to $\wg{\ell}$. In particular,
this implies that the Killing vector $\wg{\ell}$ is a null vector
on the throats:
\be \label{e:rigidity}
	\left. \wg{\ell}\cdot\wg{\ell} \right| _{{\cal S}_1} = 0 
	\qquad \mbox{and} \qquad
	\left. \wg{\ell}\cdot\wg{\ell} \right| _{{\cal S}_2} = 0 \ .  
\ee
As a guideline, note that this condition is verified by the 
helical Killing vector 
${\partial / \partial t_0} + \Omega_{\rm H} {\partial/\partial \varphi_0}$
of the Kerr spacetime, where ${\partial / \partial t_0}$, 
$\partial/\partial \varphi_0$ and $\Omega_{\rm H}$ are respectively
the Killing vector associated with stationarity, the Killing 
vector associated with axisymmetry and the rotation angular velocity
of the black hole. This classical result is known as the {\em rigidity
theorem} in the black hole literature \cite{Carte73}.

In a recent work, Friedman et al. \cite{FriedUS01} note that
corotation (in the above sense) is the only possible rotation
state consistent with the helical symmetry in the full Einstein
theory. However, a weaker definition of quasi-equilibrium (not assuming
that $\wg{\ell}$ is an exact Killing vector, as we do here) allows for
more general rotation states, as shown very recently by Cook \cite{Cook01}.

Combining Eqs.~(\ref{e:l=d/dt}) and (\ref{e:def_beta}) shows that
$\wg{\ell}$ is related to the lapse function, unit normal and shift
vector through
\be  \label{e:l_3p1}
	\wg{\ell} = N \w{n} + \wg{\beta} \ , 
\ee
so that the scalar square of $\wg{\ell}$ is
\be
	\wg{\ell}\cdot\wg{\ell} = - N^2 + \wg{\beta}\cdot\wg{\beta} \ . 
\ee
Thanks to the vanishing of the lapse on the throats, the 
rigidity condition (\ref{e:rigidity}) is then equivalent to
$\wg{\beta}\cdot\wg{\beta} = 0$ on ${\cal S}_1$ and ${\cal S}_2$. 
But $\wg{\beta}$ being a vector parallel to $\Sigma_t$, 
$\wg{\beta}\cdot\wg{\beta} = \wg{\gamma}(\wg{\beta},\wg{\beta})$;
the positive definiteness of the 3-metric $\wg{\gamma}$ implies
then
\be \label{e:shift_zero_troat}
	\left. \wg{\beta} \right| _{{\cal S}_1} = 0 
	\qquad \mbox{and} \qquad
	\left. \wg{\beta} \right| _{{\cal S}_2} = 0 \ .  
\ee
Hence, not only the $r$-component of $\wg{\beta}$ is zero 
[Eq.~(\ref{e:shift_r_zero_throat})], but the total
vector $\wg{\beta}$ vanishes on the throats.

\section{Einstein equations}	\label{s:einstein}

\subsection{General form} \label{s:gen_form}

The vacuum Einstein equations can be written \cite{York79} as
the Hamiltonian constraint equation:
\be \label{e:ham_contr}
    R - K_{ij} K^{ij} = 0 \ , 
\ee
the momentum constraint equation:
\be \label{e:mom_contr}
    D_j K^{ij} = 0 \ , 
\ee
and the ``dynamical'' equations:
\be  \label{e:dK_ij/dt}
    {\partial K_{ij}\over \partial t} - {\pounds_{\wg{\beta}}} K_{ij} 
	= - D_i D_j N + N \left( R_{ij} - 2 K_{ik} K^k_{\ j} \right) \ ,
\ee
where $R_{ij}$ denotes the Ricci tensor of the 3-metric $\wg{\gamma}$,
$R=R_i^{\ i}$ the Ricci curvature scalar, and 
$D_i$ the covariant derivative associated with $\wg{\gamma}$.
Note that we have used the vanishing of the trace of $\w{K}$, as
a consequence of the maximal slicing (assumption (3) in 
Sec.~\ref{s:metric_prop}). Besides, the geometrical relation
(\ref{e:def_extr_curv}) 
involving the extrinsic curvature results in the following equation:
\be \label{e:dg_ij/dt}
    {\partial \gamma_{ij}\over \partial t} - {\pounds_{\wg{\beta}}} \gamma_{ij} 
	= - 2 N K_{ij} \ . 
\ee

Following York \cite{York72}, Shibata \& Nakamura \cite{ShibaN95}, and
Baumgarte \& Shapiro \cite{BaumgS98}, we introduce the ``conformal
metric''
\be
	\tilde\gamma_{ij} := \gamma^{-1/3} \gamma_{ij} \ ,
\ee
where $\gamma$ is the determinant of the 3-metric components $\gamma_{ij}$.
$\tilde\gamma_{ij}$ is a tensor density of weight $-2/3$.
York \cite{York72} has shown that it carries the
dynamics of the gravitational field.
One can introduce on $\Sigma_t$ a covariant derivative $\tilde D_i$
such that
\begin{itemize}
\item[(i)] $\tilde D_i \tilde \gamma_{ij} = 0$ ;
\item[(ii)] if $\gamma_{ij}$ is conformally flat
($\gamma_{ij} = \Psi^4 f_{ij}$), then $\tilde D_i = \bar{D}_i$,
where $\bar{D}_i$ is the covariant derivative associated with
the flat metric $f_{ij}$.
\end{itemize}
We refer to Refs.~\cite{ShibaN95,BaumgS98}
for details in the case of Cartesian coordinates and to
Ref.~\cite{Gourg01} for any coordinate system. Note that the
property (i) is not sufficient to fully characterize $\tilde D_i$
since the covariant derivative $D_i$ fulfills it as well, reflecting the
fact that $\tilde \gamma_{ij}$ is a metric density and not a metric tensor:
there exists at least two distinct covariant derivatives ``associated''
with it. Let us denote by $\tilde R_{ij}$ the Ricci tensor associated
with the covariant derivative $\tilde D_i$ and by $\tilde R$ the
corresponding scalar density: $\tilde R:= \tilde \gamma^{kl} \tilde R_{kl}$,
where $\tilde \gamma^{ij}$ is the inverse conformal metric
\be 
	\tilde \gamma^{ij} := \gamma^{1/3} \gamma^{ij} \ . 
\ee
Let us also introduce the following tensor densities
\be \label{e:Aij}
	\tilde A_{ij} := \gamma^{-1/3} K_{ij} 
	\qquad \mbox{and} \qquad 
	\tilde A^{ij} := \gamma^{1/3} K^{ij} \ ,
\ee
and denote by $\tilde D^i$ the operator $\tilde \gamma^{ik}\tilde D_k$. 
The Hamiltonian constraint equation (\ref{e:ham_contr}) can then
be written as an equation for the determinant $\gamma$:
\be \label{e:ham_contr_gam}
	\tilde D_i \tilde D^i \ln\gamma
	+ {1\over 12} \tilde D_i \ln \gamma \, \tilde D^i \ln \gamma
	= {3\over 2} \left( \tilde R
		- \gamma^{1/3} \tilde A_{ij} \tilde A^{ij} \right) \ .
\ee
The momentum constraint equation (\ref{e:mom_contr}) becomes
\be \label{e:mom_contr_A}
	\tilde D_j \tilde A^{ij} + {1\over 2} \tilde A^{ij}
		\tilde D_j\ln \gamma = 0 \ .
\ee
The dynamical Einstein equations (\ref{e:dK_ij/dt}) can be
decomposed into their trace part
\be \label{e:lap_tilde_N}
	\tilde D_i \tilde D^i N + {1\over 6} \tilde D_i\ln \gamma
		\, \tilde D^i N = \gamma^{1\over 3}
				N \tilde A_{ij} \tilde A^{ij}
\ee
and their traceless part
\begin{eqnarray}
  & &  N \left( \tilde \gamma^{ik} \tilde \gamma^{jl}
      	\tilde R_{kl} + {1\over 18} \tilde D^i \ln \gamma
      		\, \tilde D^j \ln\gamma \right)
      	+ {1\over 3} \left( \tilde D^i \ln \gamma \, \tilde D^j N
      		+  \tilde D^j \ln \gamma \, \tilde D^i N \right)
	- \, \gamma^{-1/6} \tilde D^i \tilde D^j(\gamma^{1/6}N)
  					\nonumber \\
  & & - {1\over 3} \left[
	N \left( \tilde R + {1\over 18} \tilde D_k \ln\gamma
  			\, \tilde D^k\ln\gamma\right) 
   	+ {2\over 3} \tilde D_k\ln\gamma \, \tilde D^k N
    - \, \gamma^{-1/6} \tilde D_k \tilde D^k (\gamma^{1/6} N) \right] 
		\tilde \gamma^{ij}	\nonumber \\
  & & + \tilde \gamma^{1/3} \left(
  	2 N \tilde \gamma_{kl} \tilde A^{ik} \tilde A^{jl}
	+ \beta^k \tilde D_k \tilde A^{ij}
      - \tilde A^{kj} \tilde D_k \beta^i
      - \tilde A^{ik} \tilde D_k \beta^j
      + {2\over 3} \tilde D_k \beta^k \tilde A^{ij} \right)
      		= 0 \label{e:evol_aij} \ .
\end{eqnarray}
Note that in Eqs.~(\ref{e:lap_tilde_N}) and (\ref{e:evol_aij}), we have 
used the helical symmetry to set to zero the time derivatives and
that we have explicited the Lie derivatives along $\wg{\beta}$. 
Similarly, the evolution equation (\ref{e:dg_ij/dt}) for $\gamma_{ij}$
can be split into its trace part
\be  \label{e:div_tilde_beta}
	\tilde D_i \beta^i = - {1\over 2} \beta^i \tilde D_i \ln\gamma 
\ee
and its traceless part
\be \label{e:A_ij=tilde_shift}
	2 N \tilde A^{ij} = \tilde D^i\beta^j + \tilde D^j \beta^i
		- {2 \over 3} \tilde D_k \beta^k \, \tilde \gamma^{ij} \ .  
\ee
Inserting this relation into the momentum constraint (\ref{e:mom_contr_A})
results in the following equation for the shift vector:
\be \label{e:mom_constr_tilde_beta}
  \tilde D_j \tilde D^j \beta^i + {1\over 3} \tilde D^i \tilde D_j \beta^j
	+ \tilde \gamma^{ij} \tilde R_{jk} \beta^k 
   + \tilde A^{ij} \left( N \tilde D_j\ln\gamma -2 \tilde D_j N \right) = 0 \ . 
\ee
We recognize here the {\em minimal distortion} equation of Smarr \& York
\cite{SmarrY78a}, i.e. we recover the fact that the shift vector of
coordinates co-moving with respect to a Killing vector field is necessarily
a minimal distortion shift.

\subsection{Approximation of a conformally flat 3-metric}

\subsubsection{Equations} \label{s:conf_flat_eq}

As a first step in this research project, we introduce the approximation
of a conformally flat 3-metric:
\be \label{e:conf_flat}
	\wg{\gamma} = \Psi^4 \w{f} \ , 
\ee
$\Psi$ being some scalar field, and
$\w{f}$ the canonical flat 3-metric associated with the canonical
coordinates $(x_{\rm I},y_{\rm I},z_{\rm I})$ and
 $(x_{\rm II},y_{\rm II},z_{\rm II})$ (see Sec.~\ref{s:construct}).

Such an approximation has been used in all previous studies of binary
black hole initial data based on the conformal imaging approach
\cite{Cook91,CookCDKMO93,Cook94,PfeifTC00,DieneJKN00} or on the
puncture approach \cite{BrandB97,Baumg00}. It has been relaxed 
in the recently developed Kerr-Schild approach 
\cite{MatznHS99,MarroHLLMS00,MarroM00}. 
Strictly speaking, 
the assumption (\ref{e:conf_flat}) is exact only for a single non-rotating
(Schwarzschild) black hole. However, as discussed by Mathews et al. 
\cite{MatheMW98}, such an approximation is quite good even for a maximally
rotating Kerr black hole. 

As an immediate consequence of Eq.~(\ref{e:conf_flat}), we have
\be 
	\gamma = \Psi^{12} \, f \ , 
\ee
where $f$ is the determinant of the metric components $f_{ij}$. 
The conformal ``metric'' takes then the simple form
\be
	\tilde \gamma_{ij} = f^{-1/3} f_{ij} \qquad \mbox{and} \qquad
	\tilde \gamma^{ij} = f^{1/3} f^{ij}  \ .
\ee

By property (ii) of the covariant derivative $\tilde D_i$ (see
Sect.~\ref{s:gen_form} above), another consequence of Eq.~(\ref{e:conf_flat})
is that
\be \label{e:tilde_D=bar_D}
	\tilde D_i = \bar D_i \ , 
\ee
where $\bar D_i$ is the covariant derivative associated with the 
flat metric $\w{f}$. Note that, from the definition of $\tilde D^i$,
one has
\be
	\tilde D^i = f^{1/3} \bar D^i \ , 
\ee
where $\bar D^i := f^{ik} \bar D_k$. 
It follows immediately from Eq.~(\ref{e:tilde_D=bar_D}) that the
Ricci tensor $\tilde R_{ij}$ is identically zero. 

The relations (\ref{e:Aij}) can be rewritten
\be 
	\tilde A_{ij}= f^{-1/3} \hat A_{ij} 
	\qquad \mbox{and} \qquad 
	\tilde A^{ij}= f^{1/3} \hat A^{ij} \ , 
\ee
where we have introduced the tensor fields:
\be \label{e:hat_Aij}
	\hat A_{ij} := \Psi^{-4} K_{ij}  
	\qquad \mbox{and} \qquad 
	\hat A^{ij} := \Psi^4 K^{ij} \ . 
\ee

Taking into account the above relations, the Hamiltonian constraint
equation (\ref{e:ham_contr_gam}) becomes an elliptic equation for 
                                                                                       $\Psi$:
\be \label{e:eq_psi}
	\Delta \Psi = - {\Psi^5 \over 8} \hat A_{ij} \hat A^{ij} \ , 
\ee
where $\Delta := \bar D_k \bar D^k$ is the Laplacian operator with
respect to the flat metric $\w{f}$. 

The momentum constraint equation, under the form (\ref{e:mom_constr_tilde_beta}),
becomes
\be   \label{e:eq_beta}
   \Delta \beta^i + {1\over 3} \bar D^i \bar D_j \beta^j 
	= 2 \hat A^{ij} \left( \bar D_j N - 6 N \bar D_j \ln\Psi \right) \ ,
\ee
whereas the trace part of the dynamical Einstein equations,
Eq.~(\ref{e:lap_tilde_N}), becomes
\be    \label{e:eq_lapse}
   \Delta N = N \Psi^4 \hat A_{ij} \hat A^{ij}
   		- 2 \bar D_j\ln\Psi \, \bar D^j N \ .
\ee
The traceless dynamical Einstein equations (\ref{e:evol_aij})
reduces to
\begin{eqnarray}
  & & 2 N \bar D^i \ln\Psi \, \bar D^j \ln \Psi + \bar D^i \ln \Psi
  	\, \bar D^j N + \bar D^j \ln\Psi \, \bar D^i N
  	- {1\over 4} \Psi^{-2} \bar D^i \bar D^j (\Psi^2 N) \nonumber \\
  & & - {1\over 3} \left[ 2 N \bar D_k \ln\Psi \bar D^k \ln \Psi
    + 2 \bar D_k \ln\Psi \bar D^k N - {1\over 4} \Psi^{-2}
    	\Delta (\Psi^2 N) \right] f^{ij} \nonumber \\
  & & + {\Psi^4\over 4} \left[ 2 N  f_{kl} \hat A^{ik} \hat A^{jl}
  	+ \beta^k \bar D_k(\hat A^{ij})
  	- \hat A^{kj} \bar D_k \beta^i  - \hat A^{ik} \bar D_k \beta^j
  	+ {2\over 3} \bar D_k \beta^k \, \hat A^{ij} \right] = 0 \ .
  							\label{e:eq_dyn}
\end{eqnarray}

The trace part of the evolution equation for $\gamma_{ij}$,
Eq.~(\ref{e:div_tilde_beta}), becomes
\be \label{e:div_bar_beta}
	\bar D_i \beta^i = - 6 \beta^i \bar D_i \ln \Psi \ ,
\ee
whereas its traceless part (\ref{e:A_ij=tilde_shift}) results in a
relation between the extrinsic curvature tensor
and the shift vector:
\be \label{e:hat_Aij_shift}
    \hat A^{ij} = {1\over 2N} (L\beta)^{ij} \ ,
\ee
where $(L\beta)^{ij}$ denotes the (flat) 
conformal Killing operator \cite{York74}
applied to the vector $\wg{\beta}$:
\be    \label{e:conformal_Killing}
    (L\beta)^{ij} := \bar D^i \beta^j + \bar D^j \beta^i
		- {2\over 3} \bar D_k \beta^k \, f^{ij}  \ .
\ee

\subsubsection{Solution scheme} \label{s:scheme}

Our approach is the following one:
using Eq.~(\ref{e:hat_Aij_shift}) to evaluate $A^{ij}$,
consider Eqs.~(\ref{e:eq_psi}),
(\ref{e:eq_beta}) and (\ref{e:eq_lapse}) as coupled elliptic equations to be
solved for respectively $\Psi$, $\wg{\beta}$ and $N$.
The remaining five Einstein equations, Eqs.~(\ref{e:eq_dyn}), are not used to get
the solution. Moreover, they are not satisfied by the solution
$(\Psi,\wg{\beta},N)$, except in special circumstances (e.g. 
spherical symmetry).
This reflects the fact that the conformally flat form (\ref{e:conf_flat})
constitutes only an approximation to the exact Einstein equations.
An interesting application of Eqs.~(\ref{e:eq_dyn})
is then to evaluate its left-hand side in order to
gauge the error resulting from the conformally flat approximation. 
Besides, note that Eq.~(\ref{e:div_bar_beta}) is not used in the above
scheme. We will discuss this point in Sec.~\ref{s:regul}.

The system of Eqs.~(\ref{e:eq_psi})-(\ref{e:eq_lapse}), resulting
from the assumption of conformal flatness and maximal slicing, 
has been already proposed by Isenberg \& Nester \cite{IsenbN80}, 
as well as Wilson \& Mathews \cite{WilsoM89}, as an interesting
approximation to the full Einstein equations. It has been 
used by many authors to compute binary neutron stars on circular orbits
\cite{BaumgCSST97,BaumgCSST98,MarroMW98,MarroMW99,BonazGM99a,GourgGTMB01,%
UryuE00,UryuSE00}.

\subsubsection{Boundary conditions}\label{s:boundary}

The equations (\ref{e:eq_psi}), (\ref{e:eq_beta}) and (\ref{e:eq_lapse})
we are facing being elliptic, it is very important to discuss the
boundary conditions to set on their solutions. Thanks to the isometry $I$,
the computational domain is chosen to be half the full spacetime, i.e.
only ${\cal M}_{\rm I}$.
Its boundaries are then the spatial infinity and the two throats
${\cal S}_1$ and ${\cal S}_2$. At spatial infinity, the metric should
be asymptotically flat (hypothesis (1) of Sec.~\ref{s:metric_prop}).
This implies that
\be \label{e:psi_infinity}
	\Psi \rightarrow 1 \qquad \mbox{when} \quad r_1 \rightarrow
		\infty \quad \mbox{or} \quad  r_2 \rightarrow
		\infty
\ee
and
\be \label{e:lapse_infinity}
	N \rightarrow 1 \qquad \mbox{when} \quad r_1 \rightarrow
		\infty \quad \mbox{or} \quad  r_2 \rightarrow
		\infty \ .
\ee
Combining Eqs.~(\ref{e:helical}), (\ref{e:l_3p1}) and (\ref{e:lapse_infinity}),
we get the asymptotic behavior of the shift vector:
\be \label{e:shift_infinity}
	\wg{\beta} \rightarrow \Omega {\partial \over \partial \varphi_0}
		\qquad \mbox{when} \quad r_1 \rightarrow
		\infty \quad \mbox{or} \quad  r_2 \rightarrow
		\infty \ .
\ee	

The boundary conditions on the throats have been derived in 
Sec.~\ref{s:spacetime_metric}. In particular, 
Eqs.~(\ref{e:isom_grr_der}), (\ref{e:isom_gtt_der}) and 
(\ref{e:isom_gpp_der}) are equivalent to the the following condition on 
the conformal factor $\Psi$:
\be \label{e:bound_psi}
	\left. \left( {\partial \Psi\over \partial r_1} + 
		{\Psi\over 2 r_1} \right) \right| _{{\cal S}_1} = 0 
	\qquad \mbox{and} \qquad
	\left. \left( {\partial \Psi\over \partial r_2} + 
		{\Psi\over 2 r_2} \right) \right| _{{\cal S}_2} = 0 \ . 
\ee
All the remaining equations listed in 
Eqs.~(\ref{e:isom_grr_der})-(\ref{e:isom_gpp_der}) are automatically satisfied 
by  the conformally flat form (\ref{e:conf_flat}).
The boundary equation on the lapse have already been given 
[Eq.~(\ref{e:lapse_zero_throat})]:
\be \label{e:lapse_zero_throat_2}
	N|_{{\cal S}_1} = 0 \qquad \mbox{and} \qquad N|_{{\cal S}_2} = 0 \ .
\ee
as well as that on the shift vector, resulting from the rigid rotation
hypothesis [Eq.~(\ref{e:shift_zero_troat})]:
\be \label{e:shift_zero_troat_2}
	\left. \wg{\beta} \right| _{{\cal S}_1} = 0 
	\qquad \mbox{and} \qquad
	\left. \wg{\beta} \right| _{{\cal S}_2} = 0 \ .  
\ee

\subsubsection{Regularity on the throats and isometry of the shift} \label{s:regul}

A direct consequence of Eqs.~(\ref{e:lapse_zero_throat_2}) and 
(\ref{e:hat_Aij_shift}) is that the shift vector on the throats should
satisfy not only (\ref{e:shift_zero_troat_2}) but also
\be \label{e:Lbeta_zero_throat}
	\left. (L\beta)^{ij} \right| _{{\cal S}_1} = 0
	\qquad \mbox{and} \qquad
	\left. (L\beta)^{ij} \right| _{{\cal S}_2} = 0 \ ,
\ee
in order for the extrinsic curvature to be regular
on the throat. Note that in the case of a single rotating black hole,
such a condition is equivalent to 
$\partial \beta^\varphi / \partial r = 0$ and 
$\partial \beta^\varphi / \partial \theta = 0$. The first condition
follows from e.g. Eq.~(10.25) of Ref.~\cite{Carte73} and
the last one from the rigidity theorem ($\beta^\varphi$ is constant
-- and zero -- on the horizon).
In the present case, 
the properties (\ref{e:shift_r_zero_throat}) and
(\ref{e:dbr/dt_s})-(\ref{e:dbp/dr_s}), which follow from the
isometry $I$, in conjunction with the property (\ref{e:shift_zero_troat_2}),
which follows from the rigidity assumption
\footnote{Note that the constancy of $\wg{\beta}$ on the throat,
implied by Eq.~(\ref{e:shift_zero_troat_2}), results in the vanishing
of all the partial derivatives of the components $\beta^i$ with
respect to $\theta$ and $\varphi$.}, imply that 
\be \label{e:div_bar_beta_throat}
	\left. \bar D_i \beta^i  \right| _{{\cal S}} = 
	\left. {\partial \beta^r \over \partial r} \right| _{{\cal S}}	
\ee
and
\begin{eqnarray}
 & & \left. (L\beta)^{rr} \right| _{{\cal S}} = \left. {4 \over 3} 
		{\partial \beta^r \over \partial r} \right| _{{\cal S}} \\
 & & \left. (L\beta)^{\theta\theta} \right| _{{\cal S}} = \left. 
		- {2 \over 3 r^2} 
		{\partial \beta^r \over \partial r} \right| _{{\cal S}} \\
 & & \left. (L\beta)^{\varphi\varphi} \right| _{{\cal S}} = \left. 
		- {2 \over 3 r^2\sin^2\theta} 
		{\partial \beta^r \over \partial r} \right| _{{\cal S}} \\
 & & \left. (L\beta)^{r\theta} \right| _{{\cal S}} = 
	\left. (L\beta)^{r\varphi} \right| _{{\cal S}} = 
	\left. (L\beta)^{\theta\varphi} \right| _{{\cal S}} = 0 \ .  
\end{eqnarray}
Therefore the condition (\ref{e:Lbeta_zero_throat}) is equivalent to
\be \label{e:dbrdr_throat}
	\left. {\partial \beta^{r_1} \over \partial r_1} 
					\right| _{{\cal S}_1} = 0
	\qquad \mbox{and} \qquad
	\left. {\partial \beta^{r_2} \over \partial r_2} 
					\right| _{{\cal S}_2} = 0 \ . 
\ee
Now the trace of the relation between the extrinsic curvature and
the derivative of the 3-metric, Eq.~(\ref{e:div_bar_beta}), gives,
when inserting Eq.~(\ref{e:shift_zero_troat_2}) in its right-hand
side, 
\be
	\left. \bar D_k \beta^k  \right| _{{\cal S}_1} = 0 
	\qquad \mbox{and} \qquad
	\left. \bar D_k \beta^k  \right| _{{\cal S}_2} = 0 \ . 
\ee
From Eq.~(\ref{e:div_bar_beta_throat}), it follows that 
Eq.~(\ref{e:dbrdr_throat}) is satisfied as well. This establishes the
regularity property (\ref{e:Lbeta_zero_throat}). 

The above argument relies on the fact that Eq.~(\ref{e:div_bar_beta}),
relating the trace of the extrinsic curvature tensor (here zero) to the
divergence of the shift vector, is satisfied. However, as discussed in
Sec.~\ref{s:scheme}, we solve only Eqs.~(\ref{e:eq_psi})-(\ref{e:eq_lapse})
to get the metric fields $\Psi$, $N$ and $\wg{\beta}$. This means
that there is a priori no guarantee that Eq.~(\ref{e:div_bar_beta}) is
satisfied by the solution of (\ref{e:eq_psi})-(\ref{e:eq_lapse})
(see Sec.~IV.C of Ref.~\cite{FriedUS01} for a discussion of this point).
It has been argued recently by Cook~\cite{Cook01} that if one reformulates
the problem by assuming that the helical vector $\wg{\ell}$ is not an
exact Killing vector, but only an approximate one --- as it is in reality ---
then the only
freely specifiable part of the extrinsic curvature, as initial data,
is (\ref{e:hat_Aij_shift}), not (\ref{e:div_bar_beta}). This means that
the relation (\ref{e:div_bar_beta}) between the extrinsic curvature and
the shift is not as robust as the relation (\ref{e:hat_Aij_shift}).

Another problem is that, when solving the system (\ref{e:eq_psi})-(\ref{e:eq_lapse})
subject to the boundary conditions (\ref{e:bound_psi})-(\ref{e:shift_zero_troat_2}),
there is no guarantee not only that the solution for the shift vector
obeys to Eq.~(\ref{e:div_bar_beta}) but also that it obeys to the isometry
conditions  (\ref{e:dbt/dr_s}) and (\ref{e:dbp/dr_s}) [the other isometry
conditions, namely (\ref{e:shift_r_zero_throat}), (\ref{e:dbr/dt_s}) and
(\ref{e:dbr/dp_s}), are satisfied by virtue of the boundary condition
(\ref{e:shift_zero_troat_2})]. In the companion article \cite{GrandGB01},
we present a method to enforce the regularity condition (\ref{e:dbrdr_throat})
as well as the isometry condition conditions (\ref{e:dbt/dr_s}) and
(\ref{e:dbp/dr_s}). This amounts to add, at each step of the iteration,
a corrective term $\wg{\beta}_{\rm cor}$ to the solution $\wg{\beta}_{\rm constr}$
of the momentum constraint (\ref{e:eq_beta}), so that
the shift vector
\be
	\wg{\beta} = \wg{\beta}_{\rm constr} + \wg{\beta}_{\rm cor}
\ee
is well behaved, i.e. satisfies (i) the rigidity boundary conditions
(\ref{e:shift_zero_troat_2}), (ii) the condition (\ref{e:dbrdr_throat})
which ensures the regularity of the extrinsic curvature on the throats,
and (iii) the isometry condition (\ref{e:I_beta}).

If at the end of the iteration, $\wg{\beta}_{\rm cor}$ has converged to zero,
then we get a regular solution of the Einstein equations in the conformal
flatness approximation. On the contrary, if $\wg{\beta}_{\rm cor}$ stays
at some finite value, we get a solution which violates the momentum constraint
equation. The numerical solutions we have computed \cite{GrandGB01} belong to
this category. However they have (cf. Sec.~IV.B of \cite{GrandGB01})
\be
	|\wg{\beta}_{\rm cor}| < 10^{-3} |\wg{\beta}| \ ,
\ee
which shows that the momentum constraint is only very slightly violated.
Taking into account the other approximations performed, e.g. conformal flatness,
we find this to be very satisfactory.

\subsection{Global quantities}  \label{s:global}

The total mass-energy content in a $\Sigma_t$ hypersurface is given by
the Arnowitt-Deser-Misner (ADM) mass $M$, which is expressed by means of the 
surface integral at spatial infinity
\be \label{e:M_ADM_def}
    M = {1\over 16\pi} \oint_\infty f^{ik} f^{jl} \left(
	  \bar D_j \gamma_{kl} - \bar D_k \gamma_{jl} \right) \, dS_i  
\ee
(see e.g. Eq.~(20.9) of Ref.~\cite{MisneTW73}). 
In the case of the conformally flat 3-metric $\gamma_{ij} = \Psi^4 f_{ij}$, 
this integral can be written
\be \label{e:M_ADM_1}
   M = - {1\over 2\pi} \oint_\infty \bar D^i \Psi \, dS_i \ . 
\ee
By means of the Green-Ostrogradski formula, this expression can be converted
into the volume integral of $\Delta \Psi$ plus surface integrals on the
throats; using the Hamiltonian constraint (\ref{e:eq_psi})
to express $\Delta \Psi$, as well as the boundary condition
(\ref{e:bound_psi}) on the throats, one gets
\be
  M = {1\over 16\pi} \int \Psi^5 \hat A_{ij} \hat A^{ij} \, \sqrt{f} \, d^3 x 
	+ {a_1 \over 4\pi} \oint_{r_1=a_1} \!\!\!\!\! \Psi \sin\theta_1 
		\, d\theta_1 \, d\varphi_1  
	+ {a_2 \over 4\pi} \oint_{r_2=a_2} \!\!\!\!\! \Psi \sin\theta_2 
		\, d\theta_2 \, d\varphi_2  \ . 
\ee

The total angular momentum in a $\Sigma_t$ hypersurface is defined 
by the following surface integral at spatial infinity \cite{York79,York80}
\be \label{e:J_def}
  J = {1\over 8\pi}  \oint_\infty \left(
	K^i_{\ j} - K^k_{\ k} f^i_{\ j} \right) m^j \, dS_i \ , 
\ee
where $\w{m} := \partial /\partial \varphi_0$ [see Eqs.~(\ref{e:helical})
and (\ref{e:shift_infinity})] is the rotational Killing vector of the
flat metric $\w{f}$ (to which $\wg{\gamma}$ is asymptotic). Note that
in the present case $K^k_{\ k} = 0$ (maximal slicing). Note also that 
$J$ defined by Eq.~(\ref{e:J_def}) coincides with the $z$-component of the 
vector $J^i$ defined by Eq.~(12) of Bowen \& York \cite{BowenY80}. 
As discussed by York \cite{York79,York80}, 
contrary to the definition (\ref{e:M_ADM_def}) of the ADM mass, 
the definition (\ref{e:J_def}) of $J$
requires some asymptotic gauge-fixing conditions stronger than the 
mere asymptotic flatness (\ref{e:asymp_flat1})-(\ref{e:asymp_flat2}),
because of the supertranslations ambiguity. Some natural 
gauge-fixing conditions are provided by 
the asymptotic quasi-isotropic gauge proposed
by York \cite{York79}. Such conditions are fulfilled by the conformally
flat metric (\ref{e:conf_flat}). Using the fact that $\Psi=1$ at spatial
infinity, we can replace $K^i_{\ j} m^j$ by 
$\Psi^6 \hat A^{ij} f_{jk} m^k$ in Eq.~(\ref{e:J_def}) and express $J$
by means of the Green-Ostrogradski formula as
a volume integral plus surface integrals on the throats. 
The volume integral vanishes identically, as one can see by considering
the following identity:
\be
    \bar D_i (\Psi^6 \hat A^{ij} f_{jk} m^k) = \bar D_i (\Psi^6 \hat A^{ij})
	f_{jk} m^k + {1\over 2} \Psi^6 \hat A^{ij} 
	[\bar D_i( f_{jk} m^k ) + \bar D_j ( f_{ik} m^k ) ] \ .
\ee
The first term on the r.h.s. vanishes by virtue of the momentum
constraint (\ref{e:mom_contr_A}), which can be written as
\be
\label{e:mom_contr_PA}
	\bar D_i(\Psi^6 \hat A^{ij}) = 0 \ , 
\ee
whereas the second term vanishes for $\w{m}$ is a Killing vector of
the flat metric $\w{f}$. Thus the formula for $J$ reduces to 
integrals on the throats:
\be
\label{e:J_throat}
	J = - {1 \over 8\pi} \oint_{r_1=a_1} \!\!\!\!\! \Psi^6 
		\hat A^{ij} f_{jk} m^k d\bar S_i 
	     - {1 \over 8\pi} \oint_{r_2=a_2} \!\!\!\!\! \Psi^6 
		\hat A^{ij} f_{jk} m^k d\bar S_i \ , 
\ee
where $d\bar S_i$ denotes the surface element with respect to the
flat metric $\w{f}$ and oriented toward the ``interior'' of the throats.

\subsection{Determination of the orbital velocity}\label{s:section_viriel}

The orbital angular velocity $\Omega$ does not appear in the 
partial differential equations listed in Sec.~\ref{s:conf_flat_eq}.
It shows up only in the boundary condition (\ref{e:shift_infinity})
for the shift vector.
This contrasts with the binary neutron star case, where $\Omega$
enters in the equation governing the equilibrium of the fluid 
(see e.g. \cite{GourgGTMB01}). 

At this point, it appears that, solving Eqs.~(\ref{e:eq_psi})-(\ref{e:eq_lapse}),
with the boundary conditions
(\ref{e:psi_infinity})-(\ref{e:shift_infinity})
and (\ref{e:bound_psi})-(\ref{e:shift_zero_troat_2}),
one can get a solution $(N,\wg{\beta},\Psi)$
for any given value of $\Omega$. For instance, 
if we set $\Omega=0$ in the boundary condition (\ref{e:shift_infinity}),
we get $\wg{\beta}=0$ as a solution of (\ref{e:eq_beta})
and the Misner-Lindquist solution for $\Psi$ \cite{Misne63,Lindq63}.
Of course, such a solution is not admissible on physical grounds, and
we need an extra condition to fix $\Omega$. 

As a boundary condition at spatial infinity, we have demanded only
that $\w{g}$ tends to the Minkowski metric of flat spacetime
[conditions (\ref{e:asymp_flat1})-(\ref{e:asymp_flat2}) or
(\ref{e:lapse_infinity})-(\ref{e:shift_infinity})]. We could go 
a little further and demand instead that $\w{g}$ tends to the
Schwarzschild metric corresponding to the ADM mass $M$.
This implies the following behavior for the conformal factor $\Psi$
and the lapse $N$ [cf. Eqs.~(\ref{e:ds2_Schwarz}) and (\ref{e:N_Schwarz})]:
\be \label{e:psi_schwarz}
	\Psi \sim 1 + {M \over 2 r} \qquad \mbox{when} \quad r
		\rightarrow \infty \ , 
\ee
\be  \label{e:lapse_schwarz}
	N \sim 1 - {M \over r}  \qquad \mbox{when} \quad r
		\rightarrow \infty \ ,  
\ee
where $r$ denotes either the coordinate $r_1$ or $r_2$. 
From the very definition of $M$, the behavior (\ref{e:psi_schwarz})
is guaranteed by Eq.~(\ref{e:M_ADM_1}). However, the solution of (\ref{e:eq_lapse})
is such that $N\sim 1 - M'/r$, with a priori $M' \not = M$.
The behavior (\ref{e:lapse_schwarz}) is thus a extra condition imposed on the
system (\ref{e:eq_psi})-(\ref{e:eq_lapse}).
This is the condition which will enable us to fix $\Omega$.

Let us show that for stationary spacetimes, i.e. in the case
where $\wg{\ell}$ is timelike at infinity, the extra condition
(\ref{e:lapse_schwarz}) follows from the remaining Einstein equations 
(\ref{e:eq_dyn}), i.e. the equations that we have not used in the
system (\ref{e:eq_psi})-(\ref{e:eq_lapse}).
Indeed, the quadratic terms of the type $\bar D^i\ln\Psi \bar D^j N$
or $\bar D^i\ln\Psi \bar D^j \ln\Psi$ which appear in Eq.~(\ref{e:eq_dyn})
all decay at least as $r^{-4}$ when $r\rightarrow\infty$. 
Now, for stationary spacetimes,
it can be seen that the Lie derivative along $\wg{\beta}$
of $\hat A^{ij}$ which appear in Eq.~(\ref{e:eq_dyn}), decays also
at least as $r^{-4}$ [Eq.~(\ref{e:Lb_Kangu}) below].
Then Eq.~(\ref{e:eq_dyn}) implies that
$\bar D^i \bar D^j (\Psi^2 N)$ decays at least as $r^{-4}$, which 
means that the $1/r$ (monopolar) part of $\Psi^2 N$ vanishes, i.e.
\be  \label{e:viriel}
	\Psi^2 N \sim 1 + {\alpha \over r^2}  \qquad \mbox{when} \quad r
		\rightarrow \infty \ .   
\ee
This is possible only if 
$\Psi^2$ and $N$ have opposite monopolar $1/r$ terms, 
which implies the property (\ref{e:lapse_schwarz}).

Note that, for stationary spacetimes, the monopolar term of
the lapse $N$ is the Komar mass associated with the timelike
Killing vector. The  condition (\ref{e:lapse_schwarz}) is then 
intimately linked to the 
virial theorem: as already shown by two of us~\cite{GourgB94}, a
relativistic generalization of the classical virial theorem
can be obtained provided that the Komar mass coincides with the ADM
mass [property (\ref{e:lapse_schwarz})]. This last property
has been shown to hold for asymptotically
flat stationary spacetimes by Beig \cite{Beig78}.
In order to exhibit more clearly the link with the virial theorem, 
let us combine Eqs.~(\ref{e:eq_psi}) and (\ref{e:eq_lapse}) to
derive an equation for $\Psi^2 N$ (see
e.g. Eq.~(51) of Ref.~\cite{GourgGTMB01}):
\be \label{e:Psi2_N}
	\Delta (\Psi^2 N) = N \Psi^6 \left[ 4\pi  S_i^{\ i}
		+ {3\over 4} \hat A_{ij} \hat A^{ij} \right] 
		 + 2 \bar D_i \Psi \, \bar D^i(\Psi N) \ ,
\ee
where we have re-introduced a non-vanishing stress-energy tensor
$T_{\mu\nu}$ via 
$S_{\alpha\beta} := \gamma_{\alpha}^{\ \mu} \gamma_{\beta}^{\ \nu}
T_{\mu\nu}$ for the benefit of the discussion when considering
the Newtonian limit. The condition (\ref{e:viriel}) is equivalent to the
vanishing of the monopolar term of $\Psi^2 N$, i.e. from 
Eq.~(\ref{e:Psi2_N}) and assuming a spacelike slice $\Sigma_t$
diffeomorphic to $\RR^3$, 
\be  \label{e:GRV3}
	\int_{\Sigma_t} \left\{
	N \Psi^6 \left[ 4\pi  S_i^{\ i}
		+ {3\over 4} \hat A_{ij} \hat A^{ij} \right]
		 + 2 \bar D_i \Psi \, \bar D^i(\Psi N)
				\right\} \sqrt{f} \, d^3x = 0 \ .
\ee
It is easy to see that this relation is equivalent to the
relativistic virial theorem given by Eq.~(29) of Ref.~\cite{GourgB94},
once the latter is specialized to a conformally flat 3-metric.
The Newtonian limit of Eq.~(\ref{e:GRV3})
is nothing but the classical virial theorem:
\be \label{e:viriel_newt}
	2 T + 3 P + W = 0 \ , 
\ee
where $T$ is the total kinetic energy of the system, $P$ the volume 
integral of the pressure and $W$ the gravitational potential energy. 
Note that the value of $\Omega$ for two Newtonian particles
of individual mass $m$ in
circular orbit (radius $R$) can be obtained from Eq.~(\ref{e:viriel_newt})
($T=mR^2\Omega^2$, $P=0$, $W=-m^2/(2R)$); 
this results of course in the Keplerian value
$\Omega^2 = 2 m/(2R)^3$. 

Let us remark that Detweiler \cite{Detwe89}
has proposed to determine the orbital velocity $\Omega$ of 
binary black holes in circular orbits by means of a variational principle. 
Although he does not state it precisely, his variational principle 
also use the ``virial'' assumption (\ref{e:lapse_schwarz}) [cf. the not so 
well justified sentence ``In the gauge described in Chapter 19 of
Misner et al. (1973) the flux integral at infinity is 
$4\pi M - 8\pi J$'' below his Eq.~(17)].

\subsection{Generalized Smarr formula}

A formula relating the ADM mass $M$, the total angular momentum $J$, 
the angular velocity $\Omega$ and some integrals on the throats
can be obtained as follows. The key point is to notice that the
Einstein equations (\ref{e:ham_contr}), (\ref{e:mom_contr}),
the trace of (\ref{e:dK_ij/dt}) and (\ref{e:dg_ij/dt}) imply,
when $\partial/\partial t$ is a Killing vector [Eq.~(\ref{e:l=d/dt})], 
the following remarkable identity \cite{Detwe89} : 
\be \label{e:Detw}
	D^i(D_i N - K_{ij} \beta^j) = 0 \ . 
\ee
Note that this equation is fully general and does not 
assume that the 3-metric $\wg{\gamma}$ is conformally flat. 
The vanishing of the divergence (\ref{e:Detw}) enables one to use
the Green-Ostrogradski formula to get an identity involving only
surface integrals:
\be
	\oint_\infty(D_i N - K_{ij}\beta^j) dS^i 
	= - \sum_{a=1}^2 \oint_{{\cal S}_a} (D_i N - K_{ij}\beta^j) dS^i \ ,
\ee
where by convention $dS^i$ is oriented towards the ``interior''
of the throats.

From Eq.~(\ref{e:lapse_schwarz}), the flux integral of $D_i N$ on the
left hand side is equal to $4\pi M$. Using Eqs.~(\ref{e:J_def}) and
(\ref{e:shift_infinity}), the flux integral of  $K_{ij}\beta^j$
is equal to $8\pi \Omega J$. The second term on the right hand side
vanishes because $\wg{\beta}=0$ on the throats [rigidity condition, 
Eq.~(\ref{e:shift_zero_troat_2})], so that one is left with
\be
	M - 2 \Omega J = -{1\over 4\pi} \oint_{{\cal S}_1} 
					D_i N \, dS^i
			-{1\over 4\pi} \oint_{{\cal S}_2} 
					D_i N \, dS^i \ . 
\ee
This formula generalizes to the binary black hole case
the classical formula that Smarr \cite{Smarr73} derived for a 
single rotating black hole (the surface integral on the right hand side
being then the black hole surface gravity multiplied by the horizon area). 

\section{Asymptotic behavior of the fields} \label{s:asymptotic}

The asymptotic behavior (near spatial infinity) of the conformal factor
$\Psi$ and the lapse function $N$ are given by Eqs.~(\ref{e:psi_schwarz})
and (\ref{e:lapse_schwarz}).
The aim of this section is to get the asymptotic behavior of the shift
vector $\wg{\beta}$ and the extrinsic curvature tensor $\w{K}$
(or equivalently $\hat A_{ij}$).
In doing so, we will gain some insight about the assumption of asymptotic
flatness and the leftover Einstein equations (\ref{e:eq_dyn}).

\begin{figure}
\centerline{ \epsfig{figure=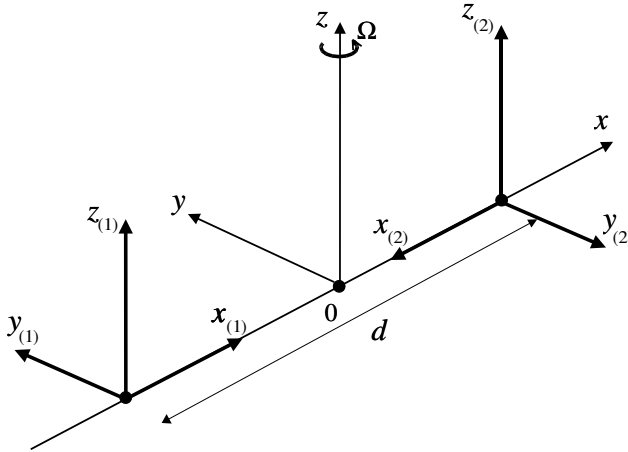,height=7cm} }
\caption[]{\label{f:coord_cart}
Cartesian coordinate systems used for the computation of the asymptotic
behavior of the shift vector.}
\end{figure}

To simplify the analysis, we restrict it to a system of identical black holes.
We introduce a Cartesian coordinate system $(x,y,z)$ such that $x$ is the
direction along the two hole centers (i.e. the centers of the spheres
$S_1$ and $S_2$), $x=0$ at the middle between the two, and $z$ is the direction
perpendicular to the orbital plane. Moreover, let us introduce coordinate
systems centered on each hole, according to
\be
	\left\{ \begin{array}{lcl}
	x_1 & = & x + d / 2\\
	y_1 & = & y \\
	z_1 & = & z \
	\end{array}      \right.
	\qquad \mbox{and} \qquad
	\left\{ \begin{array}{lcl}
	x_2 & = & - x + d / 2\\
	y_2 & = & - y \\
	z_2 & = & z
	\end{array}      \right.       \ ,
\ee
where $d$ denotes the coordinate distance between the centers of the two
spheres $S_1$ and $S_2$. These two coordinate systems are represented in
Fig.~\ref{f:coord_cart}

\subsection{Asymptotic behavior of the shift vector}

Let us split Eq.~(\ref{e:eq_beta}) for the shift vector
in two parts, assuming that its
right-hand side can be split in a part $S_1^i$ centered on hole 1 and
another part centered $S_2^i$ on hole 2. We therefore write
\be \label{e:decomp_betai}
	\beta^i = \Omega \, m^i + \beta_1^i + \beta_2^i \ ,
\ee
where $m^i$ is the rotational Killing vector $\partial/\partial\varphi$
of the flat metric $\w{f}$ already introduced in Sec.\ref{s:global},
and $\beta_1^i$ and $\beta_2^i$ are the asymptotically vanishing solutions of
\be \label{e:eq_beta_a}
   \Delta \beta^i_a + {1\over 3} \bar D^i \bar D_j \beta^j_a
	= S^i_a \ , \qquad a = 1,\ 2 \ .
\ee
Let us solve Eq.~(\ref{e:eq_beta_a}) for $a=1$
by means of the following decomposition
\cite{OoharNS87}:
\be  \label{e:shift_shibata}
	\beta^i_1 = {7\over 8} W^i_1  -{1\over 8}
		\left( {\partial \chi_1 \over \partial x^i_1}
		+ {\partial W^j_1 \over \partial x^i_1} x^j_1 \right) \ ,
\ee
where $x^i_1$ denotes the Cartesian coordinate system $(x_1,y_1,z_1)$
centered on hole 1,
and components with respect to that coordinate system
are understood in Eq.~(\ref{e:shift_shibata}).
$W^i_1$ and $\chi_1$ are solutions of the Poisson equations
\begin{eqnarray}
   \Delta W^i_1 & = & S^i_1  	\label{e:poisson_w} \\
   \Delta \chi_1 & = & - {x_1}_i S^i_1 \label{e:poisson_chi} \ .
\end{eqnarray}
Provided that the source $S^i_1$ decays at least as $r_1^{-4}$ as
$r_1\rightarrow \infty$ (\footnote{$r_1=\sqrt{x_1^2+y_1^2+z_1^2}$ is the same
radial coordinate as
that introduced in Sec.~\ref{s:construct}}), the leading behavior of the solution to
Eq.~(\ref{e:poisson_w}) is given by the harmonic function
\be  \label{e:w_monop}
	\left( W_1^{x_1}, W_1^{y_1}, W_1^{z_1} \right) =
	\left( 0, {\alpha\over r_1}, 0 \right) \ + \ O(r_1^{-2}) \ ,
\ee
where $\alpha$ is a constant. Note that we have neglected the monopolar part of
$S_1^{x_1}$ and $S_1^{z_1}$ with respect to that of $S_1^{y_1}$. This amounts
to considering that $S_1^i$ corresponds mainly to a motion along the $y_1$
direction, in accordance with the orbital motion. To understand this,
let us note that taking the Laplacian of expression (\ref{e:w_monop})
results in the following form for $S^i_1$:
\be \label{e:S_1_alpha}
	\left( S_1^{x_1}, S_1^{y_1}, S_1^{z_1} \right) =
	\left( 0, -4\pi \alpha \, \delta(x_1,y_1,z_1) , 0 \right) \ ,
\ee
where $\delta$ denotes the Dirac distribution. The Newtonian limit
for $S^i_1$ is
\be
	S_1^i = 16 \pi \rho v^i \ ,
\ee
where $\rho$ and $v^i$ denotes the matter density and velocity respectively.
This Newtonian limit holds
because in presence of matter the right-hand side of the shift equation
(\ref{e:eq_beta}) should contain the term $16\pi$ times the momentum
density of matter (see e.g. Eq.~(52) of Ref.~\cite{GourgGTMB01}).
For two point mass particles of individual mass $m$ in circular orbit
with angular velocity $\Omega$, this results in
\be
	\left( S_1^{x_1}, S_1^{y_1}, S_1^{z_1} \right) =
	\left( 0, -8\pi\, m\,d\, \Omega  \, \delta(x_1,y_1,z_1) , 0 \right) \ .
\ee
Identification with (\ref{e:S_1_alpha}) leads to the Newtonian value of
the coefficient $\alpha$:
\be \label{e:alpha_newt}
	\alpha_{\rm Newt} = 2 \, m \, d\,  \Omega \ .
\ee

Regarding the Poisson equation (\ref{e:poisson_chi}) for $\chi_1$, we notice
that its source has a vanishing effective mass, at least if its leading order
is as (\ref{e:S_1_alpha}); consequently, the solution $\chi_1$ has no monopolar
term in $r_1^{-1}$ and decays as $r_1^{-2}$. This means that its gradient
-- which enters in expression (\ref{e:shift_shibata}) for the shift vector --
decays as $r_1^{-3}$. Now, in this section, we are interested in the behavior
of the shift vector up to the order $r^{-2}$ only. Therefore, we discard
the solution for $\chi_1$, writing
\be \label{e:chi_no_monop}
	\chi_1 = O(r_1^{-2}) \ .
\ee
Inserting (\ref{e:w_monop}) and (\ref{e:chi_no_monop}) into (\ref{e:shift_shibata})
yields 
\be   \label{e:sol_beta1_lin}
	\left( \beta_1^{x_1}, \beta_1^{y_1}, \beta_1^{z_1} \right) =
	\left( {\alpha \, x_1 y_1 \over 8 r_1^3} ,
		{\alpha \over 8 r_1} \left[ 7 + {y_1^2 \over r_1^2} \right] ,
	         {\alpha \, z_1 y_1 \over 8 r_1^3}   \right) \ + \ O(r_1^{-2}) \ .
\ee

Let us remark that this solution is nothing but one of the
three (harmonic) eigenvectors of the operator $\bar D \cdot L = \Delta +
1/3 \, \bar D (\bar D\cdot )$ [cf. Eq.~(\ref{e:conformal_Killing})] which
decay as $r^{-1}$. This can be seen by comparing with the list
of these harmonic vectors provided by \'O~Murchadha \cite{Omurc92}:
the solution (\ref{e:sol_beta1_lin}) above is the item (19.2) of
\'O~Murchadha's list. Moreover, this author has shown that this harmonic
vector is generated from linear momentum in the $y_1$ direction, in full
accordance with the analysis performed above [cf. Eq.~(\ref{e:S_1_alpha})].

At this stage, our solution (\ref{e:sol_beta1_lin}) describes only the
linear momentum of hole 1. Since we are considering corotating black holes,
they must have individual angular momentum (spin), in addition to their linear
momentum, although neither the notion of individual spin nor individual linear
momentum can be defined rigorously for a binary system in general relativity
(only the total angular momentum can be defined, as in Sec.~\ref{s:global}).
To take the rotation of the black holes into account, let us add a
pure spin part to $\beta_1^i$, of the type
\be \label{e:sol_beta1_spin}
	\left( \beta_{1,\rm spin}^{x_1}, \beta_{1,\rm spin}^{y_1},
		\beta_{1,\rm spin}^{z_1} \right) =
	\left( 2 s {y_1 \over r_1^3}, -2 s {x_1\over r_1^3}, 0 \right) ,
\ee
where the constant $s$ is some parameter which measures the amount of spin,
the latter being supposed to be aligned along the $z_1$ axis.
Note that (\ref{e:sol_beta1_spin}) is a harmonic vector of the operator
$\bar D \cdot L$ which decays as $r_1^{-2}$. It is nothing but the
asymptotic part of the axisymmetric
shift vector generated by a single rotating object [see e.g. Eq.~(4.13) of
Ref.~\cite{BonazGSM93}, where $N^\varphi = - \beta_{\rm spin}^\varphi$].
Adding (\ref{e:sol_beta1_lin}) to (\ref{e:sol_beta1_spin}), we get the
following final expression for the shift vector ``mostly generated'' by hole 1:
\begin{eqnarray}
	\beta_1^{x_1}  & = & {\alpha x_1 y_1 \over 8 r_1^3} + 2 s {y_1 \over r_1^3}
			+ O(r_1^{-3}) 		\\
	\beta_1^{y_1}  & = & {\alpha \over 8 r_1}
		\left( 7 + {y_1^2 \over r_1^2} \right) -  2 s {x_1 \over r_1^3}
			+ O(r_1^{-3}) 		\\
	\beta_1^{z_1}  & = & {\alpha z_1 y_1 \over 8 r_1^3} + O(r_1^{-3})  \ . 	
\end{eqnarray}

By symmetry, we get exactly the same expression for the components of the
shift vector $\wg{\beta}_2$ with respect to the coordinates $(x_2,y_2,z_2)$.
Let us now express the components of both $\wg{\beta}_1$ and $\wg{\beta}_2$
with respect to the coordinates $(x,y,z)$ centered on the system. Taking
into account the orientations of $(x_1,y_1,z_1)$ and
$(x_2,y_2,z_2)$ with respect to $(x,y,z)$ (see Fig.~\ref{f:coord_cart}),
we obtain:
\begin{eqnarray}
	\beta_1^x  & = & {\alpha (x+d/2) y \over 8 r_1^3} + 2 s {y \over r_1^3}
			+ O(r^{-3}) 		\\
	\beta_1^y  & = & {\alpha \over 8 r_1}
		\left( 7 + {y^2 \over r_1^2} \right) -  2 s {x+d/2 \over r_1^3}
			+ O(r^{-3}) 		\\
	\beta_1^z  & = & {\alpha z y \over 8 r_1^3} + O(r^{-3}) 	
\end{eqnarray}
and
\begin{eqnarray}
	\beta_2^x  & = & {\alpha (d/2-x) y \over 8 r_2^3} + 2 s {y \over r_2^3}
			+ O(r^{-3}) 		\\
	\beta_2^y  & = & - {\alpha \over 8 r_2}
		\left( 7 + {y^2 \over r_2^2} \right) -  2 s {x-d/2 \over r_2^3}
			+ O(r^{-3}) 		\\
	\beta_2^z  & = & - {\alpha z y \over 8 r_2^3} + O(r^{-3}) 	  \ .
\end{eqnarray}
By adding together these two expressions [cf. Eq.~(\ref{e:decomp_betai})]
and performing an expansion to
the order $r^{-2}$, we get the following asymptotic form of the total
shift vector $\wg{\beta}$:
\begin{eqnarray}
	\beta^x & = & - \Omega y + {\alpha d \over 8} {y\over r^5}
		\left( -2 x^2 + y^2 + z^2 \right)
		+ 4 s {y \over r^3} + O(r^{-3}) 	\\
	\beta^y & = & \Omega x -  {\alpha d \over 8} {x\over r^5}
		\left( 7 x^2 + 10 y^2 + 7 z^2 \right)
		- 4 s {x \over r^3} + O(r^{-3}) 	\\
	\beta^z & = &  - {3 \alpha d \over 8} {x y z\over r^5}
		+ O(r^{-3}) 		\ .	
\end{eqnarray}
Note that, apart from the $\Omega$ part, the total shift decays as
$r^{-2}$, contrary to $\wg{\beta}_1$ and $\wg{\beta}_2$, which decay
as $r^{-1}$. From the above expression, 
$\wg{\beta}$ can be linearly decomposed into three parts:
\be \label{e:decomp_beta} 
	\wg{\beta} = \wg{\beta}_{\rm kin}
			+ \wg{\beta}_{\rm angu}
			+ \wg{\beta}_{\rm quad} \ ,
\ee
with
\begin{eqnarray}
	\beta_{\rm kin}^x & = & - \Omega \, y 	\\
	\beta_{\rm kin}^y & = & \Omega \, x 	\\
	\beta_{\rm kin}^z & = & 0 	\ ,	
\end{eqnarray}
\begin{eqnarray}
	\beta_{\rm angu}^x & = & \left( {\alpha d \over 2} + 4 s \right)
				  {y \over r^3} \label{e:beta_angu_x} \\
	\beta_{\rm angu}^y & = & - \left( {\alpha d \over 2} + 4 s \right)
				  {x \over r^3} \label{e:beta_angu_y} \\
	\beta_{\rm angu}^z & = & 0   \label{e:beta_angu_z}
\end{eqnarray}
and
\begin{eqnarray}
	\beta_{\rm quad}^x & = & - {3 \alpha d \over 8} {y \over r^3}
			\left( 1 + {x^2 \over r^2} \right)  
						\label{e:beta_quad_x} \\
	\beta_{\rm quad}^y & = & - {3 \alpha d \over 8} {x \over r^3}
			\left( 1 + {y^2 \over r^2} \right)  
						\label{e:beta_quad_y}  \\
	\beta_{\rm quad}^z & = & - {3 \alpha d \over 8} {x y z \over r^5}  
						\label{e:beta_quad_z} \ .
\end{eqnarray}
$\wg{\beta}_{\rm kin}$ is a pure kinematical term, which reflects 
that we use co-rotating coordinates. 
$\wg{\beta}_{\rm angu}$ is, as we will see below, the part
of the shift which carries the total angular-momentum of the system. 
As for $\wg{\beta}_{1,\rm spin}$ introduced above [Eq.~\ref{e:sol_beta1_spin})],
it has the familiar shape of a pure spin axisymmetric shift vector.

$\wg{\beta}_{\rm quad}$ is one of the nine harmonic vectors of the operator
$\bar D \cdot L$ which decay as $r^{-2}$. It has the number (22.7) in
\'O Murchadha's list \cite{Omurc92}. By the way, $\wg{\beta}_{\rm angu}$
has the number (22.1) in the same list. As shown by \'O Murchadha 
\cite{Omurc92} [cf. his Eq.~(29)],
$\wg{\beta}_{\rm quad}$ arises from the fact that the $Q_{xy}$ component
of the quadrupole moment $Q_{ij}$ of the system is time varying with respect to
some asymptotic inertial frame \footnote{In Post-Newtonian theory,
it is also well known that some part of the gravitomagnetic potential
-- the shift vector in our language -- can be generated by the first time
derivative of the mass quadrupole moment, see e.g. Sec.~VI.B of Ref.~\cite{DamouSX91}
or Eq.~(4.2) of Ref.~\cite{BrumbG01}.}.
This is the only such component.
Indeed, if we consider a Newtonian system of two identical point mass particles
on a circular orbit of diameter $d$ in the $x-y$ plane, 
the time derivative of its quadrupole moment with respect to the inertial frame
is given by 
\begin{eqnarray}
	\dot Q_{xx} & = & - m {d^2 \over 2} \Omega \sin (2 \Omega t) \\
	\dot Q_{xy} & = &  m {d^2 \over 2} \Omega \cos (2 \Omega t) 
				\label{e:dot_Qxy} \\
	\dot Q_{yy} & = &  m {d^2 \over 2} \Omega \sin (2 \Omega t) \\
	\dot Q_{xz} & = & \dot Q_{yz} = \dot Q_{zz} = 0 \ . 
\end{eqnarray}
It is clear on this expression that at time $t=0$, when the axes of the
rotating frame coincide with those of the inertial frame (our assumption
in this discussion), the only non-vanishing component is 
$\dot Q_{xy} = m d^2 \Omega / 2$. 
Combining Eqs.~(\ref{e:alpha_newt}) and (\ref{e:dot_Qxy}),
we see that the coefficient in front of the three components
(\ref{e:beta_quad_x})-(\ref{e:beta_quad_z})
of $\wg{\beta}_{\rm quad}$ is $-3/2 \, \dot Q_{xy}$. This justifies the
label {\em quad} (for quadrupole moment) given to that part of the 
shift vector. 

\subsection{Asymptotic behavior of the extrinsic curvature}

The asymptotic behavior of the extrinsic curvature tensor is deduced
from that of the shift vector via Eqs.~(\ref{e:hat_Aij}) and 
(\ref{e:hat_Aij_shift}), which allows us to write (taking into account
that both $\Psi$ and $N$ are equal to 1 at spatial infinity)
\be
	K^{ij} \sim {1\over 2} (L\beta)^{ij} \qquad \mbox{when} \quad
			r\rightarrow \infty \ . 
\ee
Since $(L\beta_{\rm kin})^{ij}=0$ for $\wg{\beta}_{\rm kin}$ is a Killing
vector of the flat metric $\w{f}$, the decomposition (\ref{e:decomp_beta})
of the shift vector leads to the following decomposition of the
extrinsic curvature tensor:
\be \label{e:decomp_K}
	\w{K} = \w{K}_{\rm angu} + \w{K}_{\rm quad} \ ,
\ee
where $K^{ij}_{\rm angu} := (L\beta_{\rm angu})^{ij}$ and
$K^{ij}_{\rm quad} := (L\beta_{\rm quad})^{ij}$. 
Inserting the formulas (\ref{e:beta_angu_x})-(\ref{e:beta_angu_z}) in the
explicit form (\ref{e:conformal_Killing}) of the operator $L$ results
in the following components of $\w{K}_{\rm angu}$:
\begin{eqnarray}
	K_{\rm angu}^{xx} & = & - 3 \left( {\alpha d \over 2} + 4 s \right)
				{x y \over r^5} \label{e:K_angu_xx} \\
	K_{\rm angu}^{xy} & = &  3 \left( {\alpha d \over 4} + 2 s \right)
				{x^2 - y^2 \over r^5} \\
	K_{\rm angu}^{xz} & = &  - 3 \left( {\alpha d \over 4} + 2 s \right)
				{y z \over r^5} \\
	K_{\rm angu}^{yy} & = &  3 \left( {\alpha d \over 2} + 4 s \right)
				{x y \over r^5} \\
	K_{\rm angu}^{yz} & = &  3 \left( {\alpha d \over 4} + 2 s \right)
				{x z \over r^5}  \\
	K_{\rm angu}^{zz} & = & 0 \ .    \label{e:K_angu_zz}
\end{eqnarray}
A similar computation for $K^{ij}_{\rm quad}$ yields
\begin{eqnarray}
	K_{\rm quad}^{xx} & = & 3 {\alpha d \over 8} {x y \over r^5}
		\left( 5 {x^2\over r^2} - 1 \right) \label{e:K_quad_xx} \\
	K_{\rm quad}^{xy} & = & 3 {\alpha d \over 8 r^5}
		\left( 5 {x^2 y^2 \over r^2} - z^2 \right) \\
	K_{\rm quad}^{xz} & = & 3 {\alpha d \over 8} {y z \over r^5}
		\left( 5 {x^2\over r^2} + 1 \right) \\
	K_{\rm quad}^{yy} & = & 3 {\alpha d \over 8} {x y \over r^5}
		\left( 5 {y^2\over r^2} - 1 \right) \\
	K_{\rm quad}^{yz} & = & 3 {\alpha d \over 8} {x z \over r^5}
		\left( 5 {y^2\over r^2} + 1 \right) \\
	K_{\rm quad}^{zz} & = & 3 {\alpha d \over 8} {x y \over r^5}
		\left( 5 {z^2\over r^2} - 3 \right) \ . \label{e:K_quad_zz}
\end{eqnarray}
Note that both $K_{\rm angu}^{ij}$ and $K_{\rm quad}^{ij}$ decay as $r^{-3}$.

If we plug the formulas (\ref{e:K_quad_xx})-(\ref{e:K_quad_zz}) 
into the surface integral (\ref{e:J_def}) which 
gives the angular momentum, we get, after some straightforward calculations 
\be
	J(\w{K}_{\rm quad}) = 0 \ . 
\ee
This means that all the angular momentum of the system is carried by
$K_{\rm angu}^{ij}$. Indeed, inserting the formulas
(\ref{e:K_angu_xx})-(\ref{e:K_angu_zz}) into Eq.~(\ref{e:J_def})
gives
\be
	J = J(\w{K}_{\rm angu}) = {\alpha d \over 4} + 2 s \ .
\ee
For a non-relativistic point mass system, $\alpha$ is given by 
Eq.~(\ref{e:alpha_newt}) so that we get
\be
	J(\w{K}_{\rm angu})_{\rm Newt} = m {d^2 \over 2} \Omega
						+ 2 s \ . 
\ee
The first term on the right-hand side is nothing but the orbital angular
momentum of the system and the second terms is the sum of the spins
$s$ of the two particles. Hence $J(\w{K}_{\rm angu})_{\rm Newt}$ is
equal to the total angular momentum of the system. 

\subsection{Helical symmetry and asymptotic flatness} \label{s:non_flat}

Let us consider the five Einstein equations that we have not taken into account
for the solution of the problem, i.e. Eqs.~(\ref{e:eq_dyn}).
Thanks to the asymptotic 
behavior (\ref{e:psi_schwarz}), (\ref{e:lapse_schwarz}) 
and (\ref{e:viriel}) of $N$ and $\Psi$, all the terms involving
products of gradients of $N$ or $\Psi$, as well as the ones involving
second derivatives of $\Psi^2 N$, decay at least as $r^{-4}$. 
The quadratic term $f_{kl}\hat A^{ik} \hat A^{jl}$ decays as $r^{-6}$
for $K^{ij}$ decays as $r^{-3}$, as seen above. 
The only remaining term in Eq.~(\ref{e:eq_dyn}) is the Lie derivative
of $\hat A^{ij}$ along $\wg{\beta}$. Asymptotically, one has
\be
	\pounds_{\wg{\beta}} \hat A^{ij} = \pounds_{\wg{\beta}} K^{ij}
						+ O(r^{-4}) \ . 
\ee
It can be seen easily that only the Lie derivative along $\wg{\beta}_{\rm kin}$
matters:
\be
	\pounds_{\wg{\beta}} K^{ij} = \pounds_{\wg{\beta}_{\rm kin}} K^{ij}
					+ O(r^{-6})    \ .
\ee
Let us introduce the splitting (\ref{e:decomp_K}) of $K^{ij}$ into
this expression. After some computations, we find that 
\be  \label{e:Lb_Kangu}
	\pounds_{\wg{\beta}} K^{ij}_{\rm angu} = O(r^{-4})  \ , 	
\ee
whereas 
\begin{eqnarray}
   \pounds_{\wg{\beta}} K^{xx}_{\rm quad} & = &
	{3 \alpha d \Omega \over 8 r^7} 
	(4 x^4 - 5 x^2 y^2 - 3 x^2 z^2 + y^4 - y^2 z^2 -2 z^4) + O(r^{-4}) \\
   \pounds_{\wg{\beta}} K^{xy}_{\rm quad} & = &
	{15 \alpha d \Omega \over 8 r^7} 
	x y (x^2 - y^2) + O(r^{-4}) \\
   \pounds_{\wg{\beta}} K^{xz}_{\rm quad} & = &
	{3 \alpha d \Omega \over 8 r^7} 
	 x z (7x^2 - 3 y^2 + 2 z^2) + O(r^{-4}) \\
   \pounds_{\wg{\beta}} K^{yy}_{\rm quad} & = &
	{3 \alpha d \Omega \over 8 r^7} 
	 (-x^4 + 5 x^2 y^2 + x^2 z^2 -4 y^4 + 3 x^2 z^2 + 2 z^4) + O(r^{-4}) \\
   \pounds_{\wg{\beta}} K^{yz}_{\rm quad} & = &
	{3 \alpha d \Omega \over 8 r^7} 
	 y z (3x^2 - 7 y^2 -2 z^2) + O(r^{-4}) \\
   \pounds_{\wg{\beta}} K^{zz}_{\rm quad} & = &
	- {3 \alpha d \Omega \over 8 r^7} 
	 (x^2 - y^2) (3x^2 + 3y^2 -2 z^2) + O(r^{-4}) \ ,
\end{eqnarray}
which means that $\pounds_{\wg{\beta}} K^{ij}_{\rm quad}$ decays only as
$r^{-3}$. We face here the incompatibility of helical symmetry 
and asymptotic flatness for systems that have a time-varying quadrupole 
moment (recall that $\wg{\beta}_{\rm quad}$ and hence $\w{K}_{\rm quad}$
is due to $\dot Q_{xy} \not = 0$).
Indeed $\pounds_{\wg{\beta}} K^{ij}_{\rm quad}$ is the only term
in the Einstein equations (\ref{e:eq_dyn}) which decays as slower as $r^{-3}$.
It therefore cannot be compensated by another term. This means that
the five Einstein equations (\ref{e:eq_dyn}) are violated.
Note that this problem does not arise from the assumption 
of conformal flatness of the 3-metric $\wg{\gamma}$. Relaxing this condition
would have resulted in asymptotic behaviors
of $\wg{\beta}$ and $\w{K}$ which would have been the same as that obtained
here.
Note also that for a system such as an isolated rotating axisymmetric star
(or more generally for any stationary system),
$\wg{\beta}_{\rm quad} = 0$ and $\w{K}_{\rm quad}=0$, so that
the problem of asymptotic flatness in Eq.~(\ref{e:eq_dyn}) does not 
arise. 

\section{Conclusions} \label{s:conclu}

We have presented an approach to the problem of binary
black holes in circular orbit which is similar to that 
previously used in the literature to treat binary neutron stars 
\cite{BaumgCSST97,BaumgCSST98,MarroMW98,MarroMW99,BonazGM99a,GourgGTMB01,%
UryuE00,UryuSE00}, namely an approach based on the existence of a 
helical Killing vector field along with the simplifying assumption of a 
conformally flat 3-metric. The differences between the two approaches lie in 
the boundary conditions on the throats in the black hole case.
We have chosen a spacetime manifold with spatial sections of the
Misner-Lindquist type, i.e. composed of two isometric asymptotically
flat sheets. Moreover, we have chosen the black holes to be corotating,
which has the simple geometrical interpretation of the throats being
Killing horizons.

By enforcing the isometry conditions on the shift vector, as well as
the equation relating the trace of the extrinsic curvature to the
divergence of the shift, possibly at the price of slightly modifying
the momentum constraint, all the quantities which enter in the equations
remain regular. Notably the extrinsic curvature tensor
remains finite on the throats, although the lapse vanishes there.

We have proposed to compute the orbital angular velocity of the system 
by requiring that the conformal factor $\Psi$ and the lapse function
$N$ have the same monopolar $1/r$ term in their asymptotic expansions. 
This requirement reduces to the classical virial theorem 
at the Newtonian limit. 

Contrary to previous numerical approaches
-- the conformal imaging one
\cite{KulkaSY83,Cook91,CookCDKMO93,Cook94,PfeifTC00,DieneJKN00}
and the puncture one \cite{BrandB97,Baumg00} --
our method amounts to solving five, and not four (the four constraints), 
of the Einstein equations. This reflects the fact that we have re-introduced
the time dimension in the problem. 

The formulation presented here has been implemented by means of a numerical code
based on a multi-domain spectral method and we present the first
results in the companion paper \cite{GrandGB01}. These  
results can be used as initial data for computing the black hole
coalescence within the 3+1 formalism 
\cite{BrandCGHLL00,AlcubBBLNST01,BakerBCLT01}. 

Let us stress that the work presented in this article constitutes a first 
attempt to tackle the problem of binary black hole in circular orbits. In order
to fully specify the problem and search for a unique solution,
we had to make a number of concrete choices 
which have some degree of arbitrariness,
such as the two-sheeted topology, the isometry across the throats and
the resulting boundary conditions, or the rigid rotation of the black holes.
These hypotheses could be changed to different ones, as for instance 
considering irrotational black holes instead of corotating ones. 
This shall be investigated in future works.

\acknowledgements
This work has benefited from numerous discussions with 
Luc Blanchet, Brandon Carter, Thibault Damour, David Hobill, 
J\'er\^ome Novak and Keisuke Taniguchi. We warmly thank all of them.

\end{document}